# Statistical tests based on Rényi entropy estimation

Mehmet Siddik Çadirci[1], Dafydd Evans[2], Nikolai Leonenko[2], Vitali Makogin[3]
and Oleg Seleznjev[4]

[1] *Faculty of Science, Department of Statistics, Cumhuriyet University, Sivas, Turkey.*
[2] *School of Mathematics, Cardiff University, Cardiff, Wales, UK.*
[3] *Institute for Stochastics, University of Ulm, Germany.*
[4] *Department of Mathematics and Mathematical Statistics, Umeå University, Umeå, Sweden.*

20 January 2025

**Abstract**

Entropy and its various generalizations are important in many fields, including mathematical statistics, communication theory, physics and computer science, for characterizing the amount of information associated with a probability distribution. In this paper we propose goodness-of-fit statistics for the multivariate Student and multivariate Pearson type II distributions, based on the maximum entropy principle and a class of estimators for Rényi entropy based on nearest neighbour distances. We prove the $L^2$-consistency of these statistics using results on the subadditivity of Euclidean functionals on nearest neighbour graphs, and investigate their rate of convergence and asymptotic distribution using Monte Carlo methods.

## 1 Introduction

Entropy is a measure of randomness that emerged from information theory, and its estimation plays an important role in many fields including mathematical statistics, cryptography, machine learning and indeed almost every branch of science and engineering. There are many possible definitions of entropy, for example, the *differential* or *Shannon entropy* of a multivariate density function $f : \mathbb{R}^m \to \mathbb{R}$ is defined by

$$H_1(f) = -\int_{\mathbb{R}^m} f(x) \log f(x) \, dx. \tag{1}$$

In this paper, we propose statistical tests for a class of multivariate Student and Pearson type II distribtuions, based on estimation of their Rényi entropy

$$H_q(f) = \frac{1}{1-q} \log \int_{\mathbb{R}^m} f^q(x) \, dx, \quad q \neq 1. \tag{2}$$

Estimation of Shannon and Rényi entropies for absolutely continuous multivariate distributions has been considered by many authors, including Kozachenko and Leonenko (1987), Goria *et al.* (2005), Evans (2008), Leonenko *et al.* (2008), Leonenko and Pronzato (2010), Penrose and Yukich (2011), Delattre and Fournier (2017), Gao *et al.* (2018), Bulinski and Dimitrov (2019), Berrett *et al.* (2019), Leonenko *et al.* (2021) and Ryu *et al.* (2022).

The quadratic Rényi entropy was investigated by Leonenko and Seleznjev (2010). An entropy-based goodness-of-fit test for generalized Gaussian distributions is presented by Cadirci *et al.* (2022). A recent application to image processing can be found in Dresvyanskiy *et al.* (2020).

The remainder of this paper is organized as follows. In Section 2, we present maximum entropy principles for Rényi entropy. In Section 3, we provide nearest-neighbour estimators for Rényi entropy. In Section 4, we propose statistical tests for the multivariate Student and Pearson II distributions. In Section 5, we report the results of numerical experiments.

## 2 Maximum entropy principles

Let $X \in \mathbb{R}^m$ be a random vector that has a density function $f(x)$ with respect to Lebesgue measure on $\mathbb{R}^m$, and let $S = \{x \in \mathbb{R}^m : f(x) > 0\}$ be the support of the distribution. The Rényi entropy of order $q \in (0, 1) \cup (1, \infty)$

of the distribution is

$$H_q(f) = \frac{1}{1-q} \log \int_S f^q(x)\,dx, \tag{3}$$

which is continuous and non-increasing in $q$. If the support has finite Lebesgue measure $|S|$, then

$$\lim_{q \to 0} H_q(f) = \log |S|,$$

otherwise $H_q(f) \to \infty$ as $q \to 0$. Note also that

$$\lim_{q \to 1} H_q(f) = H_1(f) = -\int_S f(x) \log f(x)\,dx.$$

Let $a \in \mathbb{R}^m$ and let $\Sigma$ be a symmetric positive definite $m \times m$ matrix.

- The *multivariate Gaussian* distribution $N_m(a, \Sigma)$ on $\mathbb{R}^m$ has density function

$$f_{a,\Sigma}^G(x) = (2\pi)^{-m/2} |\Sigma|^{-1/2} \exp\left(-\frac{1}{2}(x-a)' \Sigma^{-1} (x-a)\right).$$

For $X \sim N_m(a, \Sigma)$, we have $a = \mathbb{E}(X)$ and $\Sigma = \text{Cov}(X)$, where $\text{Cov}(X) = \mathbb{E}[(X-a)(X-a)']$ is the covariance matrix of the distribution.

- For $\nu > 0$, the *multivariate Student distribution* $T_m(a, \Sigma, \nu)$ on $\mathbb{R}^m$ has density function

$$f_{a,\Sigma,\nu}^S(x) = c_1 |\Sigma|^{-1/2} \left(1 + \frac{1}{\nu}(x-a)'\Sigma^{-1}(x-a)\right)^{-\frac{\nu+m}{2}} \quad \text{where } c_1(m,\nu) = \frac{\Gamma[(\nu+m)/2]}{(\pi\nu)^{m/2}\Gamma(\nu/2)}. \tag{4}$$

For $X \sim T_m(a, \Sigma, \nu)$ we have $a = \mathbb{E}(X)$ when $\nu > 1$ and $\Sigma = (1 - 2/\nu)\text{Cov}(X)$ when $\nu > 2$, see (Johnson and Vignat 2007). It is known that $f_{a,\Sigma,\nu}^S(x) \to f_{a,\Sigma}^G(x)$ as $\nu \to \infty$.

- For $\eta > 0$, the *multivariate Pearson Type II* distribution $P_m(a, \Sigma, \eta)$ on $\mathbb{R}^m$, also known as the *Barenblatt* distribution, has density function

$$f_{a,\Sigma,\eta}^P(x) = c_1^* |\Sigma|^{-1/2} \left[1 - (x-a)'\Sigma^{-1}(x-a)\right]_+^\eta \quad \text{where } c_1^*(m,\eta) = \frac{\Gamma(m/2 + \eta + 1)}{\pi^{m/2}\Gamma(\eta+1)}. \tag{5}$$

and $t_+ = \max\{t, 0\}$. For $X \sim P_m(a, \Sigma, \eta)$ we have $a = \mathbb{E}(X)$ and $\Sigma = (m + 2\eta + 2)\text{Cov}(X)$. It is known that $f_{a,\Sigma,\eta}^P(x) \to f_{a,\Sigma}^G(x)$ as $\eta \to \infty$.

**Remark 1.** If the covariance matrix $C$ is diagonal, the Pearson Type II distribution belongs to the class of time-dependent distributions

$$u(x,t) = c(\beta, \eta) t^{-\alpha m} \left(1 - \left(\frac{\|x\|}{ct^\alpha}\right)^\beta\right)_+^\eta$$

with $c > 0$, $\text{supp}\{u(x,t)\} = \{x \in \mathbb{R}^m : \|x\| \leq ct^\alpha\}$ and

$$c(\beta, \eta) = \beta \eta \left(\frac{m}{2}\right) \Big/ \left[2c^m \pi^{\frac{m}{2}} B\left(\frac{m}{\beta}, \eta + 1\right)\right],$$

which are known as *Barenblatt solutions* of the source-type non-linear diffusion equations $u_t' = \Delta(u^q)$, where $q > 1$, $\Delta$ is the Laplacian and $\eta = 1/(q-1)$. For details, see Frank (2005), Vázquez (2007), and De Gregorio and Garra (2020).

## 2.1 Rényi entropy

The Rényi entropy of the multivariate Gaussian distribution $N_m(a, \Sigma)$ is

$$H_q(f_{a,\Sigma}^G) = \log\left[(2\pi)^{m/2}|\Sigma|^{1/2}\right] - \frac{m}{2(1-q)} \log q = H_1(f_{a,\Sigma}^G) - \frac{m}{2}\left(1 + \frac{\log q}{1-q}\right)$$

where $H_1(f_{a,\Sigma}^G) = \log\left[(2\pi e)^{m/2}|\Sigma|^{1/2}\right]$ is the differential entropy of $N_m(a, \Sigma)$. From Zografos and Nadarajah (2005), the Rényi entropy of the multivariate Student distribution $T_m(a, \Sigma, \nu)$ is

$$H_q(f_{a,\Sigma,\nu}^S) = \frac{1}{2}\log|\Sigma| + c_2(m,\nu,q) \tag{6}$$

where

$$c_2(m,\nu,q) = \frac{1}{1-q} \log \left( \frac{B\left(q\left(\frac{\nu+m}{2}\right) - \frac{m}{2}, \frac{m}{2}\right)}{B\left(\frac{\nu}{2}, \frac{m}{2}\right)^q} \right) + \frac{m}{2} \log(\pi\nu) - \log \Gamma\left(\frac{m}{2}\right).$$

Likewise, the Rényi entropy of the multivariate Pearson Type II distribution $P_m(a, \Sigma, \eta)$ is

$$H_q(f^P_{a,\Sigma,\eta}) = \frac{1}{2} \log |\Sigma| + c_2^*(m, \eta, q), \tag{7}$$

where

$$c_2^*(m, \eta, q) = \frac{1}{1-q} \log \left( \frac{B\left(q\eta + 1, \frac{m}{2}\right)}{B\left(\eta + 1, \frac{m}{2}\right)^q} \right) + \frac{m}{2} \log(\pi) - \log \Gamma\left(\frac{m}{2}\right).$$

## 2.2 Maximum entropy principle

**Definition 2.** Let $\mathcal{K}$ be the class of density functions supported on $\mathbb{R}^m$, and subject to the constraints

$$\int_{\mathbb{R}^m} x f(x)\, dx = a \quad \text{and} \quad \int_{\mathbb{R}^m} (x-a)(x-a)' f(x)\, dx = C$$

where $a \in \mathbb{R}^m$ and $C$ is a symmetric and positive definite $m \times m$ matrix.

It is well-known that the differential entropy $H_1$ is uniquely maximized by the multivariate normal distribution $N_m(a, \Sigma)$ with $\Sigma = C$, that is

$$H_1(f) \leq H_1(f^G_{a,\Sigma}) = \log \left[ (2\pi e)^{m/2} |\Sigma|^{1/2} \right]$$

with equality if and only if $f = f^G_{a,\Sigma}$ almost everywhere. The following result is discussed by Kotz and Nadarajah (2004), Lutwak *et al.* (2004), Heyde and Leonenko (2005), and Johnson and Vignat (2007).

**Theorem 3** (Maximum Rényi entropy).

(1) For $m/(m+2) < q < 1$, $H_q(f)$ is uniquely maximized over $\mathcal{K}$ by the multivariate Student distribution $T_m(a, \Sigma, \nu)$ with $\Sigma = (1 - 2/\nu)C$ and $\nu = 2/(1-q) - m$.

(2) For $q > 1$, $H_q(f)$ is uniquely maximized over $\mathcal{K}$ by the multivariate Pearson Type II distribution $P_m(a, \Sigma, \eta)$ with $\Sigma = (2\eta + m + 2)C$ and $\eta = 1/(q-1)$.

Applying (6) and (7) yields the following expressions for the maximum entropy.

**Corollary 4.**

(1) For $m/(m+2) < q < 1$ the maximum value of $H_q$ is

$$H_q^{\max} = \frac{1}{2} \log |\Sigma| + c_2(m, q, \nu)$$

with $\Sigma = (1 - 2/\nu)C$ and $\nu = 2/(1-q) - m$.

(2) For $q > 1$ the maximum value of $H_q$ is

$$H_q^{\max} = \frac{1}{2} \log |\Sigma| + c_2^*(m, \eta, q)$$

with $\Sigma = (2\eta + m + 2)C$ and $\eta = 1/(q-1)$.

## 3 Statistical estimation of Rényi entropy

We state some known results on the statistical estimation of Rényi entropy due to Leonenko *et al.* (2008), and Penrose and Yukich (2011). Extensions of these results can be found in Penrose and Yukich (2003), Berrett *et al.* (2019), Delattre and Fournier (2017), Bulinski and Dimitrov (2019), and Gao *et al.* (2018). Let $X \in \mathbb{R}^m$ be a random vector with density function $f$, and let $G_q(f)$ denote the expected value of $f^{q-1}(X)$,

$$G_q(f) = \mathbb{E}\left[f^{q-1}(X)\right] = \int_{\mathbb{R}^m} f^q(x)\, dx, \quad q \neq 1,$$

so that $H_q(f) = \frac{1}{1-q} \log G_q(f)$.

Let $X_1, X_2, \ldots, X_N$ be independent random vectors from the distribution of $X$, and for $k \in \mathbb{N}$ with $k < N$, let $\rho_{N,k,i}$ denote the *k-nearest neighbour distance* of $X_i$ among the points $X_1, X_2, \ldots, X_N$, defined to be the $k$th order statistic of the $N-1$ distances $\|X_i - X_j\|$ with $j \neq i$,

$$\rho_{N,1,i} \leq \rho_{N,2,i} \leq \cdots \leq \rho_{N,N-1,i}.$$

We estimate the expectation $G_q(f) = \mathbb{E}(f^{q-1})$ by the sample mean

$$\hat{G}_{N,k,q} = \frac{1}{N} \sum_{i=1}^{N} (\zeta_{N,k,i})^{1-q},$$

where

$$\zeta_{N,k,i} = (N-1) C_k V_m \rho_{N,k,i}^m \quad \text{with} \quad C_k = \left[\frac{\Gamma(k)}{\Gamma(k+1-q)}\right]^{\frac{1}{1-q}}$$

and $V_m = \frac{\pi^{\frac{m}{2}}}{\Gamma(\frac{m}{2}+1)}$ is the volume of the unit ball in $\mathbb{R}^m$. The $k$-nearest neighbour estimator of $H_q$ for $q \neq 1$ is then defined to be

$$\hat{H}_{N,k,q} = \frac{1}{1-q} \log \hat{G}_{N,k,q}$$

provided that $k > q - 1$, and for the Shannon entropy $H_1$ by $\hat{H}_{N,k,1} = \lim_{q \to 1} \hat{H}_{N,k,q}$.

**Definition 5.** For $r > 0$, the *r-moment* of a density function $f$ is

$$M_r(f) = \mathbb{E}(\|X\|^r) = \int_{\mathbb{R}^m} \|x\|^r f(x) \, dx,$$

and the *critical moment* of $f$ is

$$r_c(f) = \sup\{r > 0 : M_r(f) < \infty\}$$

so that $M_r(f) < \infty$ if and only if $r < r_c(f)$.

The following result was stated without proof in Leonenko and Pronzato (2010): here we present the proof.

**Theorem 6.** Let $0 < q < 1$ and $k \geq 1$ be fixed.

1. If $G_q(f) < \infty$ and

$$r_c(f) > \frac{m(1-q)}{q}, \tag{8}$$

then $\mathbb{E}\left[\hat{G}_{k,N,q}\right] \to G_q(f)$ as $N \to \infty$. \hfill (9)

2. If $G_q(f) < \infty$, $q > \frac{1}{2}$ and

$$r_c(f) > \frac{2m(1-q)}{2q-1}, \tag{10}$$

then $\mathbb{E}\left[\hat{G}_{k,N,q} - G_q(f)\right]^2 \to 0$ as $N \to \infty$. \hfill (11)

**Remark 7.** If $G_q(f) < \infty$ for $q \in \left(1, \frac{k+1}{2}\right)$ then by (Leonenko *et al.* 2008),

$$\mathbb{E}\left[\hat{G}_{k,N,q}\right] \to G_q(f) \text{ and } \mathbb{E}\left[\hat{G}_{k,N,q} - G_q(f)\right]^2 \to 0 \text{ as } N \to \infty.$$

**Remark 8.** If $G_q(f) < \infty$ for $q \in (0,1)$ and $f(x) = O(\|x\|^{-\beta})$ as $\|x\| \to \infty$ for some $\beta > m$, then $r_c(f) = \beta - m$ and condition (8) is automatically satisfied: see Penrose and Yukich (2011) for a discussion, and counterexamples showing that conditions (8) and (10) cannot be omitted in general.

*Proof of Theorem 6.* Let us write

$$\hat{G}_{k,N,q} = \frac{1}{N} \sum_{i=1}^{N} \left[(N-1)^{1/m} (C_k V_k)^{1/m} \rho_{i,k,N}\right]^{(1-q)m}.$$

We show that the method proposed by Penrose and Yukich (2013) for $k = 1$ in fact works for any fixed $k \geq 1$. By Theorem 2.1 of (Penrose and Yukich 2013), the uniform integrability condition

$$\sup_N \mathbb{E}\left[\{((N-1)(C_k V_k)\rho_{i,k,N-1}^m\}^{(1-q)p}\right] < \infty \qquad (12)$$

for some $p > 1$ (statement 1) or some $p > 2$ (statement 2) ensures the $L_p$ convergence of $\hat{G}_{k,N,q}$ to $I_q$ as $N \to \infty$. Because we only need to obtain a bound on left-hand side of (12), we can use results on the subadditivity of Euclidean functionals defined on the nearest-neighbors graph (Yukich 1998). We use the following result (Lemma 3.3) from Penrose and Yukich (2011), see also (Yukich 1998, p.85).

**Lemma 9.** Let $0 < s < m$. If $r_c(f) > \frac{ms}{m-s}$, then

$$\sum_{j=1}^{\infty} 2^{js}\left[P(A_j)\right]^{\frac{m-s}{m}} < \infty \quad \text{where} \quad P(A_j) = \int_{A_j} f(x)\,dx$$

$$\text{and } A_j = \mathcal{B}(0, 2^{j+1}) \setminus \mathcal{B}(0, 2^j) \text{ for } j = 1, 2, \ldots$$

with $\mathcal{B}(0, R) = \{x \in \mathbb{R}^m : \|x\| \leq R\}$ and $A_0 = \mathcal{B}(0, 2)$.

We continue the proof of Theorem 6. Let $b = (1-q)mp$, and note that we can always choose $p$ to ensure that $0 < 1 - b/m < 1$. By exchangeability,

$$\mathbb{E}\left[(N-1)^{1/m}(C_k V_m)^{1/m}\rho_{i,k,N-1}\right]^b$$
$$= \mathbb{E}\left(\frac{1}{N}\sum_{i=1}^N \left[(N-1)^{1/m}(C_k V_m)^{1/m}\rho_{i,k,N-1}\right]^b\right)$$
$$= \frac{(N-1)^{b/m}}{N}(C_k V_m)^{b/m}\mathbb{E}\left(\sum_{i=1}^N \rho_{i,k,N-1}^b\right)$$
$$\leq (C_k V_m)^{b/m}(N-1)^{b/m-1}\mathbb{E}(\mathcal{L}_k^b(\mathcal{X}_N)),$$

where $\mathcal{X}_N = \{X_1, X_2, \ldots, X_N\}$, and for any finite point set $\mathcal{X} \subset \mathbb{R}^m$ and $b > 0$ we write

$$\mathcal{L}_k^b(\mathcal{X}) = \sum_{x \in \mathcal{X}} \mathcal{D}_k^b(x, \mathcal{X}),$$

where $\mathcal{D}_k^b(x, \mathcal{X})$ denotes the Euclidean distance from $x$ to its $k$-nearest neighbour in the point set $\mathcal{X} \setminus \{x\}$ when $\text{card}(\mathcal{X}) \geq k$; set $\mathcal{D}_k^b(x, \mathcal{X}) = 0$ if $\text{card}(\mathcal{X}) \leq k$. The function $\mathcal{X} \mapsto \mathcal{L}_k^b(\mathcal{X})$ satisfies the subadditivity relation

$$\mathcal{L}_k^b(\mathcal{X} \cap \mathcal{Y}) \leq \mathcal{L}_k^b(\mathcal{X}) + \mathcal{L}_k^b(\mathcal{Y}) + U_k t^b \qquad (13)$$

for all $t > 0$ and finite $\mathcal{X}$ and $\mathcal{Y}$ contained in $[0, t]^m$, where $U_k = 2km^{b/2}$, $b > 0$. Indeed, if $\mathcal{X}$ has more than $k$ elements, the $k$-nearest neighbour distances of points in $\mathcal{X}$ can only become smaller when we add some other set $\mathcal{Y}$. Hence, (13) holds with $U_k = 0$ if $\mathcal{X}$ and $\mathcal{Y}$ have more than $k$ elements. If $\mathcal{X}$ has $k$ elements or fewer, then $\mathcal{L}_k^b(\mathcal{X})$ is zero, but when we add the set $\mathcal{Y}$, we gain at most $k$ new edges from points in $\mathcal{X}$ in the nearest neighbours graph, and each of these is of length most $t\sqrt{m}$ (for more details, see Yukich (1998, pp 101-103)).

Let $s(N)$ be the largest $j \in N$ such that the set $\mathcal{X}_N = \{X_1, X_2, \ldots, X_N\} \cap A_j$ is not empty. Using ideas from Yukich (1998, p.87) we have that

$$\mathcal{X}_N \cap \left(\bigcup_{j=0}^{s(N)} A_j\right) = \bigcup_{j=0}^{s(N)} (X_N \cap A_j),$$

and by the subadditivity property,

$$\mathcal{L}_k^b(\mathcal{X}_N) \leq \mathcal{L}_k^b\{X_N \cap A_{s(N)}\}$$
$$+ \mathcal{L}_k^b\left(\mathcal{X}_N \cap \left\{\bigcup_{j=0}^{s(N)-1} A_j\right\}\right) + U_k 2^{(s(N)+1)b}.$$

Applying subadditivity in the same way to the second term on the right yields

$$\mathcal{L}_k^b\left(\mathcal{X}_N \cap \left\{\bigcup_{j=0}^{s(N)-1} A_j\right\}\right) \leq \mathcal{L}_k^b(\mathcal{X}_N \cap A_{s(N)-1})$$
$$+ \mathcal{L}_k^b\left(\mathcal{X}_N \cap \left\{\bigcup_{j=0}^{s(N)-2} A_j\right\}\right) + U_k\left(2^{s(N)}\right)^b.$$

Repeatedly applying subadditivity, we arrive at

$$\mathcal{L}_k^b(X_1,\ldots,X_N) \leq \sum_{j=0}^{s(N)} \mathcal{L}_k^b(\mathcal{X}_N \cap A_j) + 2^{b+bs(N)} \frac{U_k}{1-2^{-b}}$$
$$\leq \sum_{j=0}^{s(N)} \mathcal{L}_k^b(\mathcal{X}_N \cap A_j) + 2^{bs(N)} M_k$$
$$\leq \sum_{j=0}^{s(N)} \mathcal{L}_k^b(\mathcal{X}_N \cap A_j) + M_k \max_{1\leq i \leq N} \|X_i\|^b \quad (14)$$

for some constant $M_k$ depending on $m$, $k$ and $b$. From (13) and (14), we get

$$\mathbb{E}\left((N-1)^{1/m}(C_k V_m)^{1/m} \rho_{i,k,N-1}\right)^b$$
$$\leq (C_k V_m)^{b/m}(N-1)^{b/m-1} \mathbb{E}\left(\sum_{j=0}^{s(N)} \mathcal{L}_k^b(\mathcal{X}_N \cap A_j)\right)$$
$$+ W_k \mathbb{E}\left((N-1)^{b/m-1} \max_{1\leq i\leq N} \|X_i\|^b\right) \quad (15)$$

for some constant $W_k$ depending on $m$, $k$ and $b$. Using Lemma 3.3 of Yukich (1998) we have

$$L_k^b(\mathcal{X}) \leq L_0 (\text{diam}\mathcal{X})^b (\text{card}\mathcal{X})^{1-b/m} \quad (16)$$

for some constant $L_0 > 0$. Following Penrose and Yukich (2011), by Jensen's inequality and the fact that $\text{diam}(A_j) = 2^j$, we obtain from (15) and (16) that

$$(N-1)^{b/m-1} \mathbb{E}\left(\sum_{j=0}^{s(N)} L_k^b(\mathcal{X}_N \cap A_j)\right) \leq L_1 \sum_{j=0}^{s(N)} 2^{jb} \left[\mathbb{P}(X_1 \in A_j)\right]^{1-b/m} \quad (17)$$

where $L_1 > 0$ is a constant.

Recall our assumptions that $0 < \alpha < m/\ell$ where $\ell \in \{1,2\}$ and $\alpha = (1-q)m$, and also that $r_c(f) > (\ell m \alpha)/(m - \ell\alpha)$. Setting $s = b$ in Lemma 9, we see that the left hand side of (17) is finite, so the first term on the right hand side of (15) is bounded by a constant which is independent of $N$. For a non-negative random variable $Z > 0$, we know that

$$\mathbb{E}(Z) = \int_0^\infty \mathbb{P}(Z > z)\,dz,$$

so the second term in (15) is bounded by

$$W_k \int_0^\infty \mathbb{P}\left(\max_{1\leq i\leq N} \|X_i\|^b > u \cdot N^{1-b/m}\right) du$$
$$\leq W_k \left[1 + N \int_1^\infty \mathbb{P}\left(\|X_1\|^b > \left(u^{m/(m-b)} N\right)^{1-b/m}\right) du\right] \quad (18)$$

By the Markov inequality $\mathbb{P}(Z > a) \leq \frac{1}{a}\mathbb{E}|Z|$ for $a > 0$, we get for $u \geq 1$ that

$$\mathbb{P}\left(\|X_1\|^b > \left(u^{m/(m-b)} N\right)^{1-b/m}\right)$$
$$= \mathbb{P}\left(\|X_1\|^{mb/(m-b)} > u^{m/(m-b)} N\right)$$
$$\leq \mathbb{E}\|X_1\|^{mb/(m-b)} \frac{1}{u^{m/(m-b)} N}. \quad (19)$$

From (18) and (19), we see that the second term in (15) is bounded by

$$W_k \left[ 1 + \int_1^\infty \mathbb{E}\|X_1\|^{mb/(m-b)} \frac{1}{u^{m/(m-b)}} \, du \right]$$

which is independent of $N$, because we can choose $p$ to ensure that $0 < 1 - b/m < 1$, and

$$\mathbb{E}\|X_1\|^{\frac{mp(1-q)}{1-p(1-q)}} < \infty, \text{ or equivalently } r_c(f) > \frac{mp(1-q)}{1-p(1-q)},$$

which is consistent with conditions of Theorem 2.1. Note that the function $h(p,q) = \frac{mp(1-q)}{1-p(1-q)}$ is such that $h(1,q)$ gives the right-hand side of (8) and $h(2,q)$ gives the right-hand side of (10). Moreover, if $r_c(f) > h(1,q)$ for some $q < 1$ (resp. $r_c(f) > h(2,q)$ for some $q$ satisfying $1/2 < q < 1$), we also have $r_c(f) > h(p,q)$ for some $p > 1$ (resp. $r_c(f) > h(p,q)$ for $p > 2$).

## 4 Hypothesis tests

We now restrict the class $\mathcal{K}$ to only those distributions which satisfy the following conditions: for any fixed $k \geq 1$ and $q > 1/2$,

$$\mathbb{E}(\hat{H}_{N,k,q}) \to H_q \quad \text{as } N \to \infty, \text{ and}$$

$$\hat{H}_{N,k,q} \to H_q \quad \text{in probability as } N \to \infty.$$

By Theorem 6, we know that $\mathcal{K}$ contains $T_m(a, \Sigma, \nu)$ for all $\nu > 2$ and $P_m(a, \Sigma, \eta)$ for all $\eta > 0$.

Let $X_1, X_2, \ldots, X_N$ be independent and identically distributed random vectors with common density $f \in \mathcal{K}$, and let $\hat{C}_N$ be the sample covariance matrix,

$$\hat{C}_N = \frac{1}{N-1} \sum_{i=1}^{N} (X_i - \bar{X})(X_i - \bar{X})'.$$

1. To test the hypothesis $X \sim T_m(a, \Sigma, \nu)$ where $\nu > 2$ is unknown, we define the test statistic

$$W_{N,k}(m, \nu) = H_q^{\max} - \hat{H}_{N,k,q} \tag{20}$$

where $H_q^{\max} = \frac{1}{2} \log |\hat{\Sigma}_N| + c_2(m, \nu, q)$ with $q = 1 - 2/(\nu + m)$ and $\hat{\Sigma}_N = (1 - 2/\nu)\hat{C}_N$.

2. To test the hypothesis $X \sim P_m(a, \Sigma, \eta)$ where $\eta > 0$ is unknown, we define the test statistic

$$\tilde{W}_{N,k}(m, \eta) = H_q^{\max} - \hat{H}_{N,k,q}, \tag{21}$$

where $H_q^{\max} = \frac{1}{2} \log |\hat{\Sigma}_N| + c_2^*(m, \eta, q)$ with $q = 1 + 1/\eta$ and $\hat{\Sigma}_N = (2\eta + m + 2)\hat{C}_N$.

By the law of a large numbers, $\hat{C}_N \to C$ in probability as $N \to \infty$. Thus, by Slutsky's theorem, for any fixed $k \geq 1$ we have that

$$\lim_{N \to \infty} W_{N,k}(m, \nu) \xrightarrow{p} \begin{cases} 0 & \text{if } X \sim T_m(a, \Sigma, \nu), \\ w > 0 & \text{otherwise,} \end{cases} \tag{22}$$

and

$$\lim_{N \to \infty} \tilde{W}_{N,k}(m, \eta) \xrightarrow{p} \begin{cases} 0 & \text{if } X \sim P_m(a, \Sigma, \eta), \\ w > 0 & \text{otherwise,} \end{cases} \tag{23}$$

where "$\xrightarrow{p}$" denotes convergence in probability, and $w$ is a constant that depends on the distribution of $X$.

The distributions of the statistics $W_{N,k}(m, \nu)$ and $\tilde{W}_{N,k}(m, \eta)$ are unknown. An analytical derivation of these distributions appears difficult, because the random variables $\hat{H}_{N,k}$ and $\hat{C}_N$ are not independent, and their covariance appears to be intractable, despite the fact that the asymptotic distribution of $\hat{H}_{N,k}$ can be revealed by applying the results of Chatterjee (2008), Penrose and Yukich (2011), Delattre and Fournier (2017) and Berrett *et al.* (2019), and the asymptotic distribution of $\hat{C}_N$ by the delta method. In the next section, we investigate the asymptotic behaviour of $W_{N,k}(m, \nu)$ and $\tilde{W}_{N,k}(m, \eta)$ using Monte Carlo methods.

# 5 Numerical experiments

## 5.1 Random samples

Following Johnson 1987, random samples from the $T_m(a, \Sigma, \nu)$ and $P_m(a, \Sigma, \eta)$ distributions can be generated according to the stochastic representation
$$X = RBU + a,$$
where $R$ represents the radial distance $\left[(X-a)'\Sigma^{-1}(X-a)\right]^{1/2}$, $B$ is an $m \times m$ matrix with $B^T B = \Sigma$, and $U$ is uniformly distributed on the surface of the unit sphere in $\mathbb{R}^m$. In particular,

$$
\begin{aligned}
R^2 &\sim \text{InvGamma}(m/2, m/2) & \text{yields } X &\sim T_m(a, \Sigma, \nu), \\
R^2 &\sim \text{Beta}(m/2, \eta+1) & \text{yields } X &\sim P_m(a, \Sigma, \eta).
\end{aligned}
$$

It is interesting to note that for $X \sim T_m(a, \Sigma, \nu)$ the distribution of the radial distance is independent of $\nu$ (the dependence on $\nu$ is entirely through matrix $B$).

We investigate the standardized distributions
$$
\begin{aligned}
T_m(\nu) &:= T_m(0, I_m, \nu) \text{ for } \nu > 2, \\
P_m(\eta) &:= P_m(0, I_m, \eta) \text{ for } \eta > 1.
\end{aligned}
$$
where $I_m$ be the $m \times m$ identity matrix. For notational convenience, the multivariate Gaussian distribution will be denoted by $T_m(\infty)$ or $P_m(\infty)$, the limit of the multivariate Student distribution $T_m(\nu)$ as $\nu \to \infty$, and the Pearson II distribution $P_m(\eta)$ as $\eta \to \infty$, respectively.

## 5.2 Numerical experiments for $W_{N,k}(m, \nu)$

Let $X \sim T_m(\nu)$ where $\nu > 2$ is unknown. For $\nu_0 > 2$ fixed, we test the null hypothesis that $\nu = \nu_0$ against the alternative $\nu \neq \nu_0$, based on the value of the statistic $W_{N.k}(m, \nu_0)$. By (22), large values indicate that the null hypothesis should be rejected (this is an upper-tail test).

For each combination of parameter values $m \in \{1, 2, 3\}$ and $\nu, \nu_0 \in \{3, 4, \ldots, 20, \infty\}$, where the last of these corresponds to the multivariate Gaussian distribution, we generate a random sample of size $N \in \{100, 200, \ldots, 5000\}$ from the $T_m(\nu)$ distribution, and in each case compute the empirical value of $W_{N,k}(m, \nu_0)$ for $k \in \{1, 2, 3, 4\}$. The process is repeated independently $M = 1000$ times, which yields a sample realisation $\{w_1, w_2, \ldots, w_M\}$ from the distribution of $W_{N,k}(m, \nu_0)$ on samples from $T_m(\nu)$.

We use these sample realisations to investigate

- the consistency of $W_{N,k}(m, \nu_0)$ under the null hypothesis $\nu = \nu_0$;

- the convergence of $W_{N,k}(m, \nu_0)$ on samples from $T_m(\nu)$ where $\nu \neq \nu_0$;

- the empirical distribution of $W_{N,k}(m, \nu_0)$ on samples from $T_m(\nu)$;

- the critical values of $W_{N,k}(m, \nu_0)$ when $\nu = \nu_0$ at significance level $\alpha \in \{0.01, 0.05, 0.1\}$;

- the statistical power of of $W_{N,k}(m, \nu_0)$ to detect $\nu \neq \nu_0$ at significance level $\alpha \in \{0.01, 0.05, 0.1\}$.

**Summary of results**

- $W_{N,k}(m, \nu_0)$ on samples from $T_m(\nu)$ appears to onverge to zero when $\nu = \nu_0$, and to a strictly positive constant when $\nu \neq \nu_0$, as the sample size $N \to \infty$, which verifies (22).

- The rate at which $W_{N,k}(m, \nu_0)$ converges appears to decreases with the dimension $m$. Under the null hypothesis $\nu = \nu_0$, the rate appears to increase as $\nu$ increases. Under the alternative hypothesis $\nu \neq \nu_0$, the rate appears to decrease as the difference $|\nu - \nu_0|$ increases.

- The statistical power of the associated hypothesis test to detect $\nu \neq \nu_0$ appears to increase as $|\nu - \nu_0|$ increases. The power to detect $\nu \neq \nu_0$ when $\nu$ is large and $\nu_0$ is small is relatively high, but relatively low when $\nu$ is small and $\nu_0$ is large. This is perhaps because in the latter case, $T_m(\nu)$ becomes increasingly heavy-tailed as $\nu$ decreases, leading to outliers in samples from the distribution of $W_{N,k}(m, \nu_0)$.

### 5.2.1 Consistency

Figure 1 shows the asymptotic behaviour of the test statistic $W_{N,k}(m,\nu)$ for $m \in \{1,2,3\}$ and $k \in \{1,2,3,4\}$ on samples drawn from the $T_m(\nu)$ distribution with $\nu \in \{3,4,5,10,\infty\}$, as the sample size $N \to \infty$. In each plot, the lines represent the sample mean $\bar{w}_{N,k}(m,\nu_0)$ of our observations.

Figure 1: Consistency of $W_{N,k}(m,\nu)$ for $k \in \{1,2,3,4\}$.

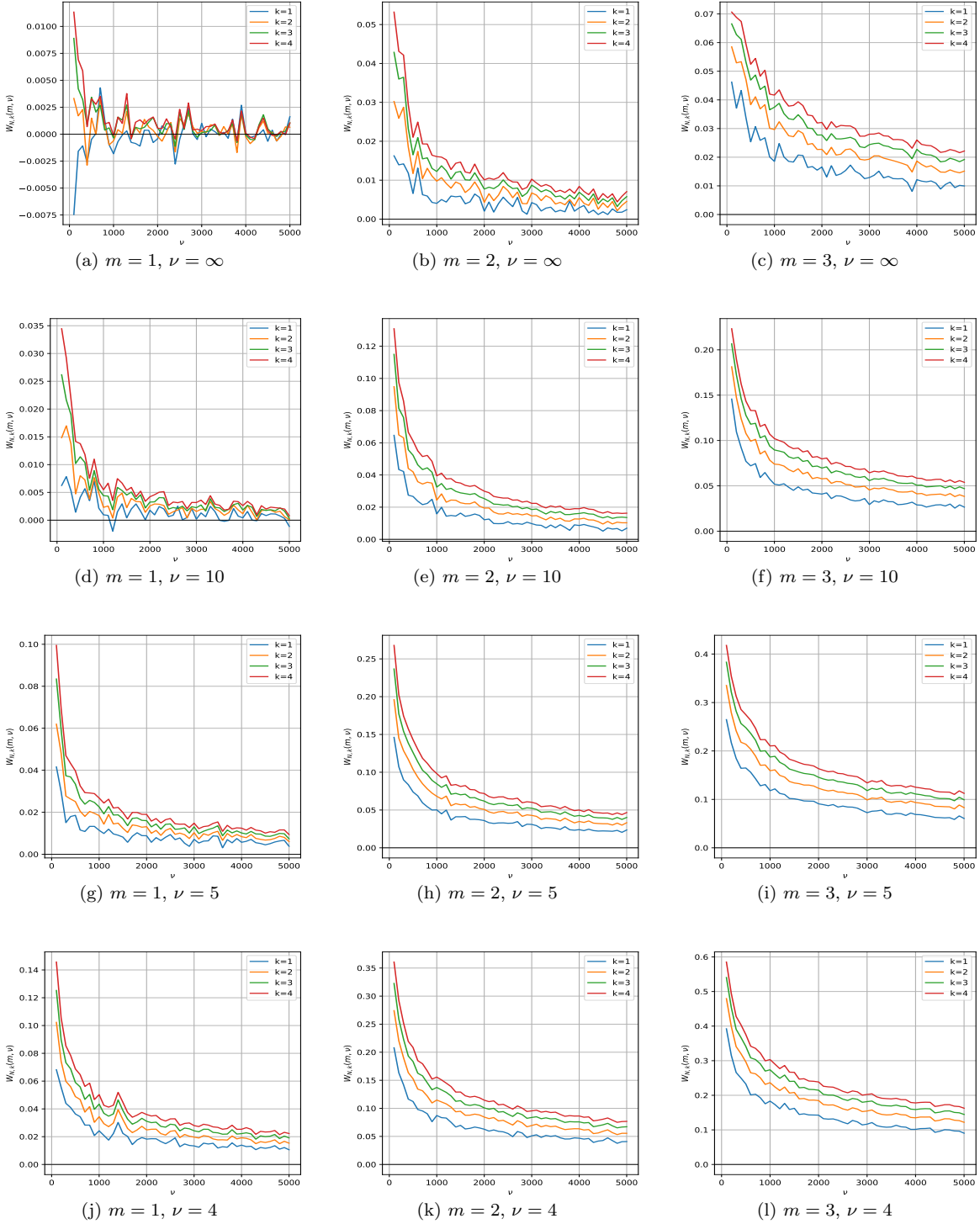

(a) $m=1, \nu=\infty$  (b) $m=2, \nu=\infty$  (c) $m=3, \nu=\infty$
(d) $m=1, \nu=10$  (e) $m=2, \nu=10$  (f) $m=3, \nu=10$
(g) $m=1, \nu=5$  (h) $m=2, \nu=5$  (i) $m=3, \nu=5$
(j) $m=1, \nu=4$  (k) $m=2, \nu=4$  (l) $m=3, \nu=4$

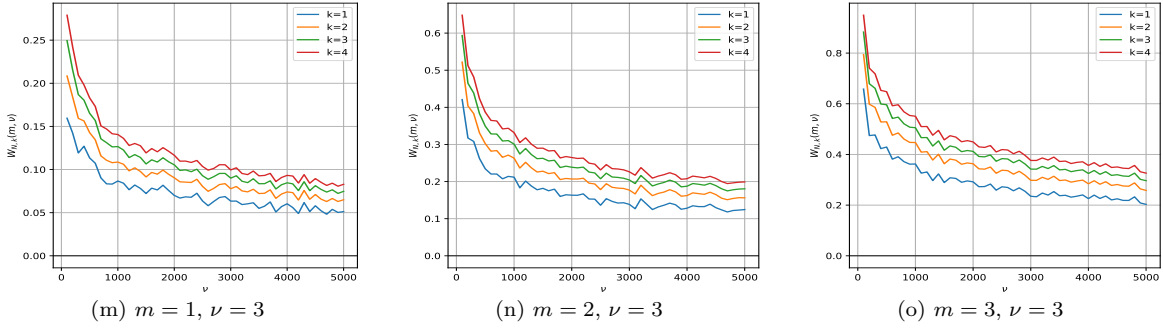

(m) $m = 1, \nu = 3$      (n) $m = 2, \nu = 3$      (o) $m = 3, \nu = 3$

Figure 1 indicates that the rate at which $W_{N,k}(m, \nu)$ convergence to zero under the null hypothesis increases as the $\nu$ increases, and decreases as the dimension $m$ increases. In addition, the mean mean value of $T_m(\nu)$ increases as the nearest neighbour index $k$ increases, but their asymptotic behaviour is otherwise very similar. For this reason, in the following we present results only for the case $k = 3$.

### 5.2.2 Convergence

Figure 2 shows the asymptotic behaviour of the test statistic $W_{N,k}(m, \nu_0)$ for $m \in \{1, 2, 3\}$ and $k = 3$ on samples drawn from the $T_m(\nu)$ distribution, with $\nu, \nu_0 \in \{3, 4, 5, 10, \infty\}$, as the sample size $N \to \infty$. In each plot, the lines represent the sample mean $\bar{w}_{N,k}(m, \nu_0)$ of the corresponding random sample, and the length of the error bars are equal to the estimated standard error $s_{N,k}(m, \nu_0)$. In all cases, numerical results indicate that $s_{N,k}(m, \nu_0) \sim N^{-1/2}$ as $N \to \infty$.

Figure 2: Convergence of $W_{N,k}(m, \nu)$ on samples from $T_m(\nu)$ for $k = 3$.

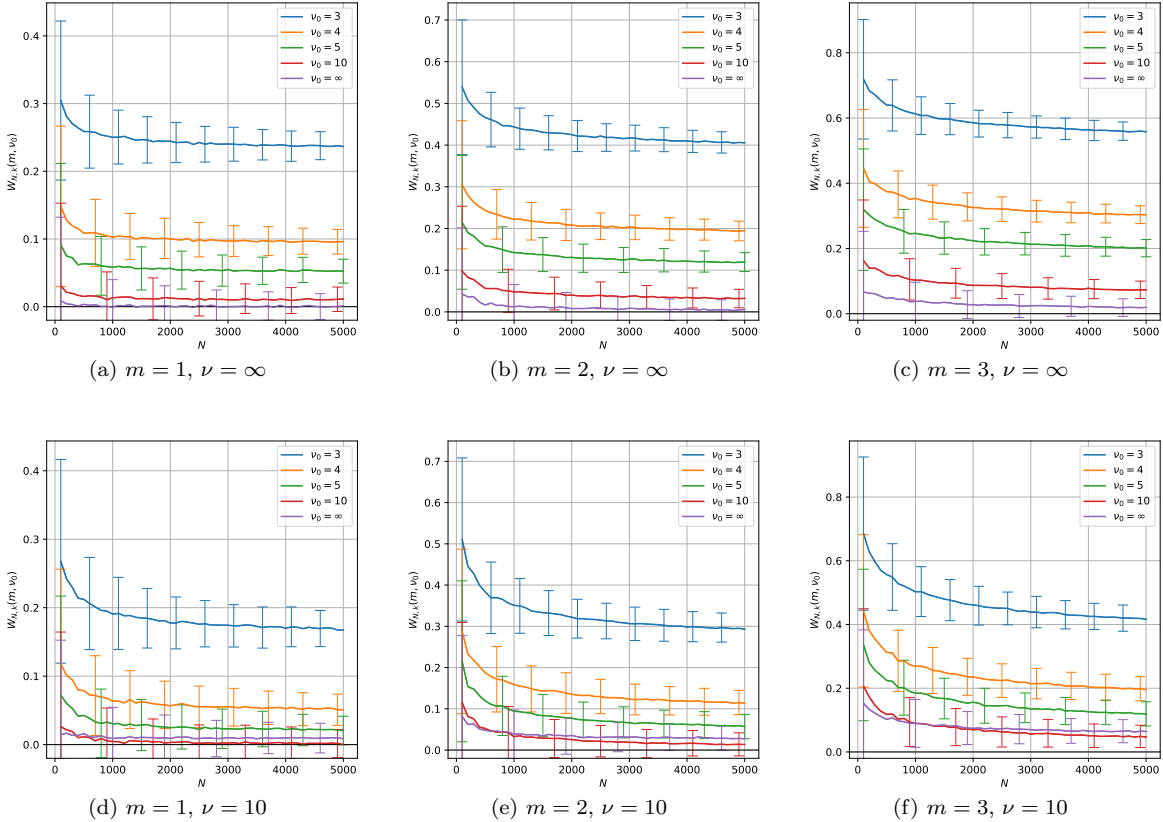

(a) $m = 1, \nu = \infty$      (b) $m = 2, \nu = \infty$      (c) $m = 3, \nu = \infty$

(d) $m = 1, \nu = 10$      (e) $m = 2, \nu = 10$      (f) $m = 3, \nu = 10$

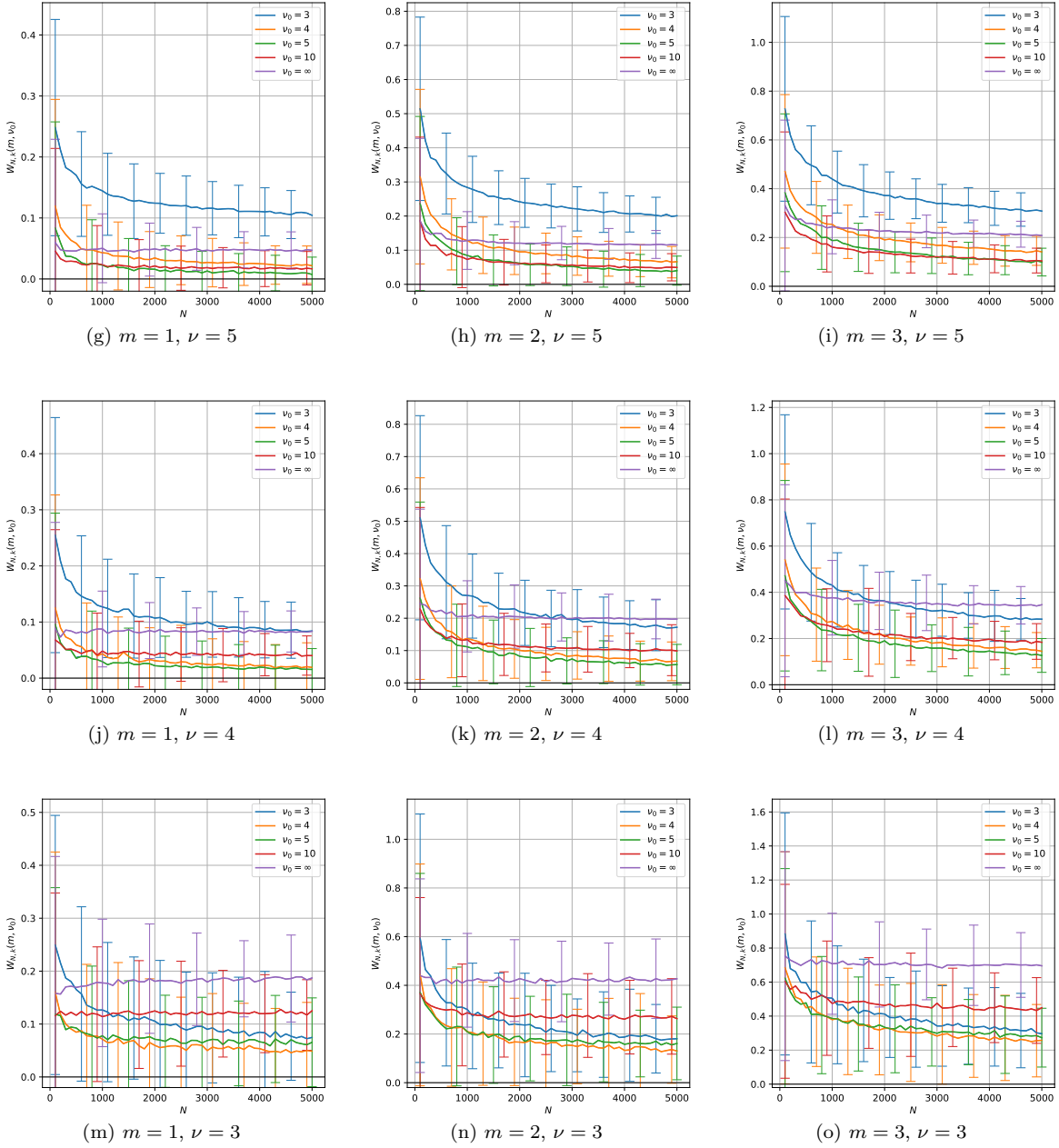

The first row of plots in Figure 2 shows the asymptotic behaviour of the test statistic $W_{N,k}(m,\nu_0)$ on samples from the multivariate Gaussian distribution $T_m(\infty)$. Here, we see that the statistic converges to zero under the null hypothesis $\nu_0 = \infty$ (purple), and to non-zero values when $\nu_0 \neq \infty$, which verifies (22). We also observe that the limiting values increase as the null distribution becomes increasingly non-Gaussian. The curve for $\nu_0 = 10$ and $\nu_0 = \infty$ are very close relative to the size of the error bars, indicating that the test will struggle to detect $\nu = \infty$ under the null hypothesis $\nu_0 = 10$ on samples of size $N = 5000$.

The second row of plots shows the asymptotic behaviour of $W_{N,k}(m,\nu_0)$ on samples from the $T_m(10)$ distribution. This time we see that $W_{N,k}(m,\nu_0)$ converges to zero when $\nu_0 = 10$ (red), and to non-zero values for $\nu_0 \neq 10$, which again verifies (22). The error bars indicate that for $N = 5000$, the test should be able to detect $\nu = 10$ under the null hypotheses $\nu_0 = 3$ and $\nu_0 = 4$ reasonably well, but not when $\nu_0 = 5$ (green) and $\nu_0 = 10$ (red).

The third and fourth row of plots show the asymptotic behaviour of $W_{N,k}(m,\nu_0)$ on samples from the $T_m(5)$ and $T_m(4)$ distributions, respectively. Here we see that for relatively large samples, the test should be able to detect $\nu = 5$ and $\nu = 4$ reasonably wellunder the null hypotheses $\nu_0 = 3$ and $\nu_0 = \infty$, but not when $\nu_0 = 10$.

The final row of plots shows the asymptotic behaviour of $W_{N,k}(m,\nu_0)$ on samples from the $T_m(3)$ distribution. Here we see that for relatively large samples, the test should be able to detect $\nu = 3$ reasonably well under the null hypotheses $\nu_0 = 10$ and $\nu_0 = \infty$, but not when $\nu_0 = 4$ and $\nu_0 = 5$.

### 5.2.3 Empirical densities

Figure 3 shows the empirical density functions of the test statistic $W_{N,k}(m,\nu_0)$ on data from the $T(\nu)$ distribution, for $\nu,\nu_0 \in \{3,4,5,10,\infty\}$ with $m \in \{1,2,3\}$, $N = 5000$ and $k = 3$.

Figure 3: Empirical density of $W_{N,k}(m,\nu_0)$ on samples from $T_m(\nu)$ for $N = 5000$ and $k = 3$.

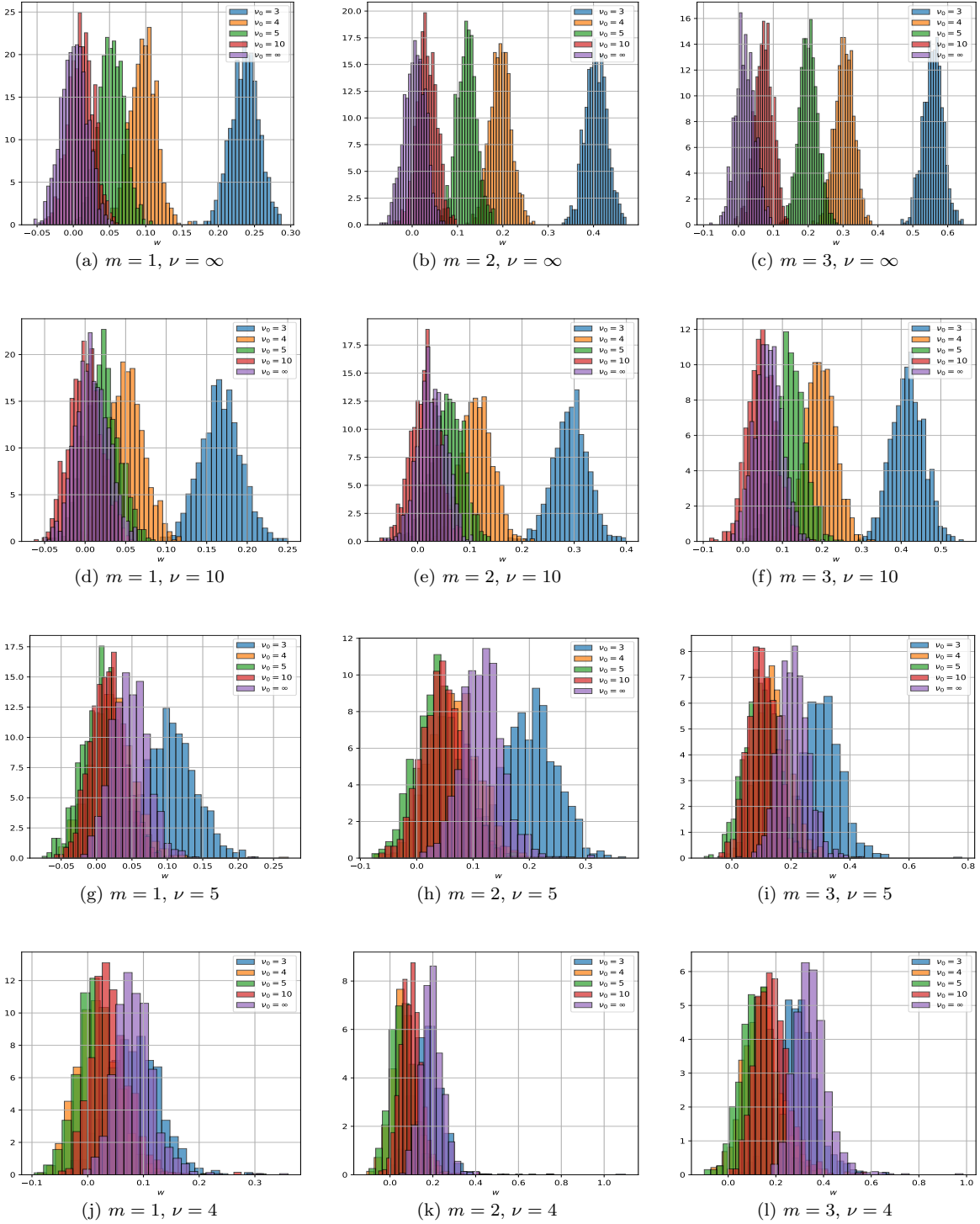

(a) $m=1, \nu=\infty$  (b) $m=2, \nu=\infty$  (c) $m=3, \nu=\infty$

(d) $m=1, \nu=10$  (e) $m=2, \nu=10$  (f) $m=3, \nu=10$

(g) $m=1, \nu=5$  (h) $m=2, \nu=5$  (i) $m=3, \nu=5$

(j) $m=1, \nu=4$  (k) $m=2, \nu=4$  (l) $m=3, \nu=4$

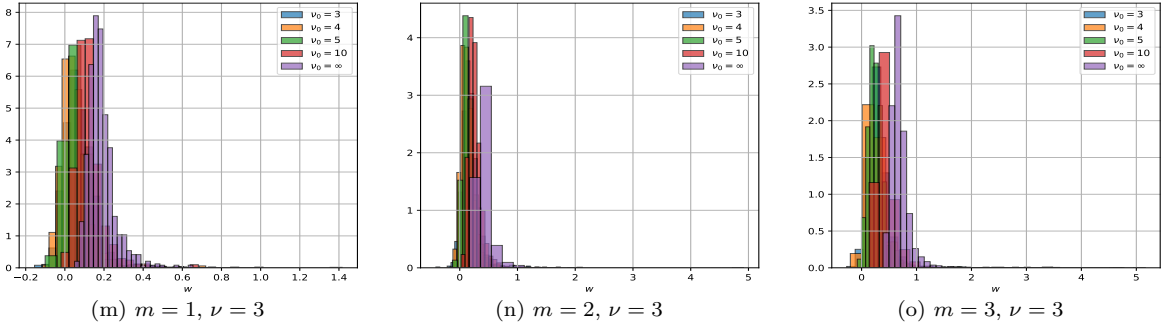

(m) $m = 1$, $\nu = 3$     (n) $m = 2$, $\nu = 3$     (o) $m = 3$, $\nu = 3$

The first row of plots in Figure 3 shows the empirical densities of $W_{N,k}(m, \nu_0)$ on samples from the multivariate Gaussian distribution $T_m(\infty)$. The distributions appear to have similar variances, with the sample mean being smallest for $\nu_0 = \infty$ (purple), which corresponds to the null hypothesis, and increasing as $\nu_0$ decreases. The empirical densities are evidently well-separated, with the separation increasing as the dimension $m$ increases, which indicates that the test should detect the alternative hypothesis $\nu \neq \infty$ with increasing probability as the test value $\nu_0$ decreases.

The second and third rows show the empirical densities of $W_{N,k}(m, \nu_0)$ on samples from the $T_m(10)$ and $T_m(5)$ distributions, respectively. We see that the separation between the empirical densities for the various values of $\nu_0$ decreases as $\nu$ decreases, indicating a decrease of statistical power to detect the alternative $\nu \neq \nu_0$ compared to the case $\nu = \infty$.

The fourth and final rows show the empirical densities on samples from the $T_m(4)$ and $T_m(3)$ distributions, respectively. Here we see a further decrease in statistical power to detect the alternative $\nu \neq \nu_0$. In addition, for these cases we note the presence of outliers in the sample of values for $W_{N,k}(m, \nu_0)$, due to the heavy-tailed nature of $T_m(4)$ and $T_m(3)$.

Figure 4 shows box plots corresponding to the empirical density functions presented in Figure 3. Here we again see a decrease in the separation between the empirical densities and an increase in the presence of outliers as $\nu$ decreases, which corresponds to $T_m(\nu)$ becoming increasingly heavy-tailed.

Figure 4: Box plots of $W_{N,k}(m, \nu_0)$ on samples from $T_m(\nu)$ for $N = 5000$ and $k = 3$.

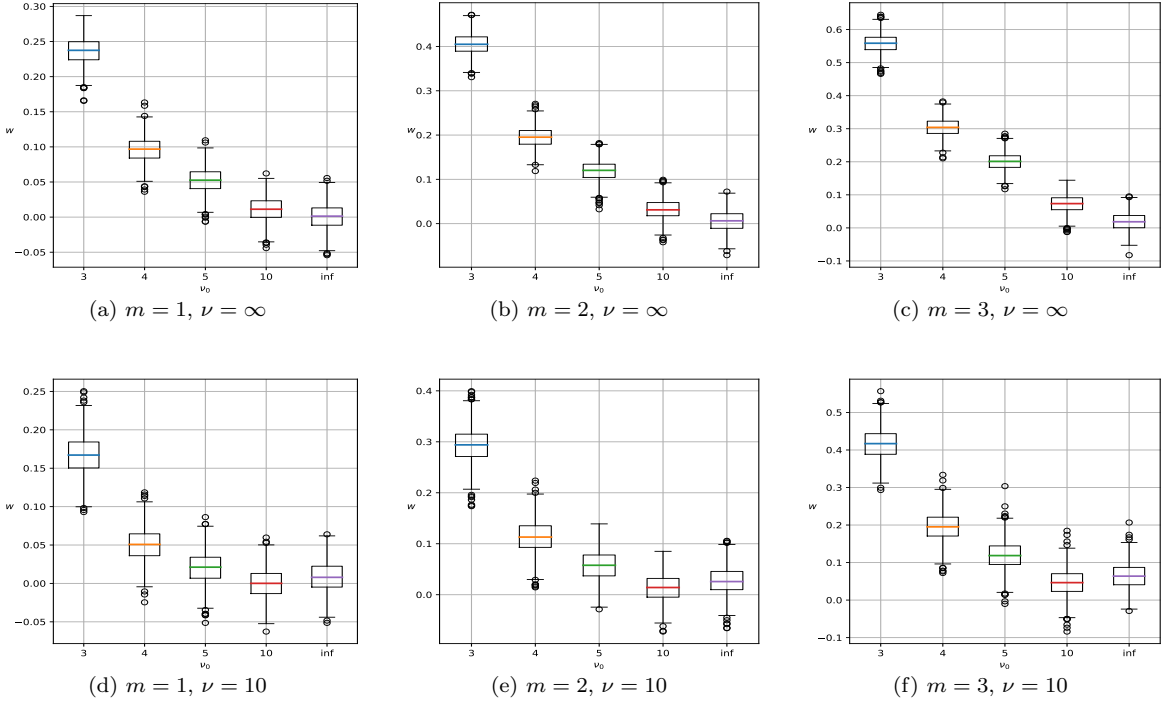

(a) $m = 1$, $\nu = \infty$     (b) $m = 2$, $\nu = \infty$     (c) $m = 3$, $\nu = \infty$

(d) $m = 1$, $\nu = 10$     (e) $m = 2$, $\nu = 10$     (f) $m = 3$, $\nu = 10$

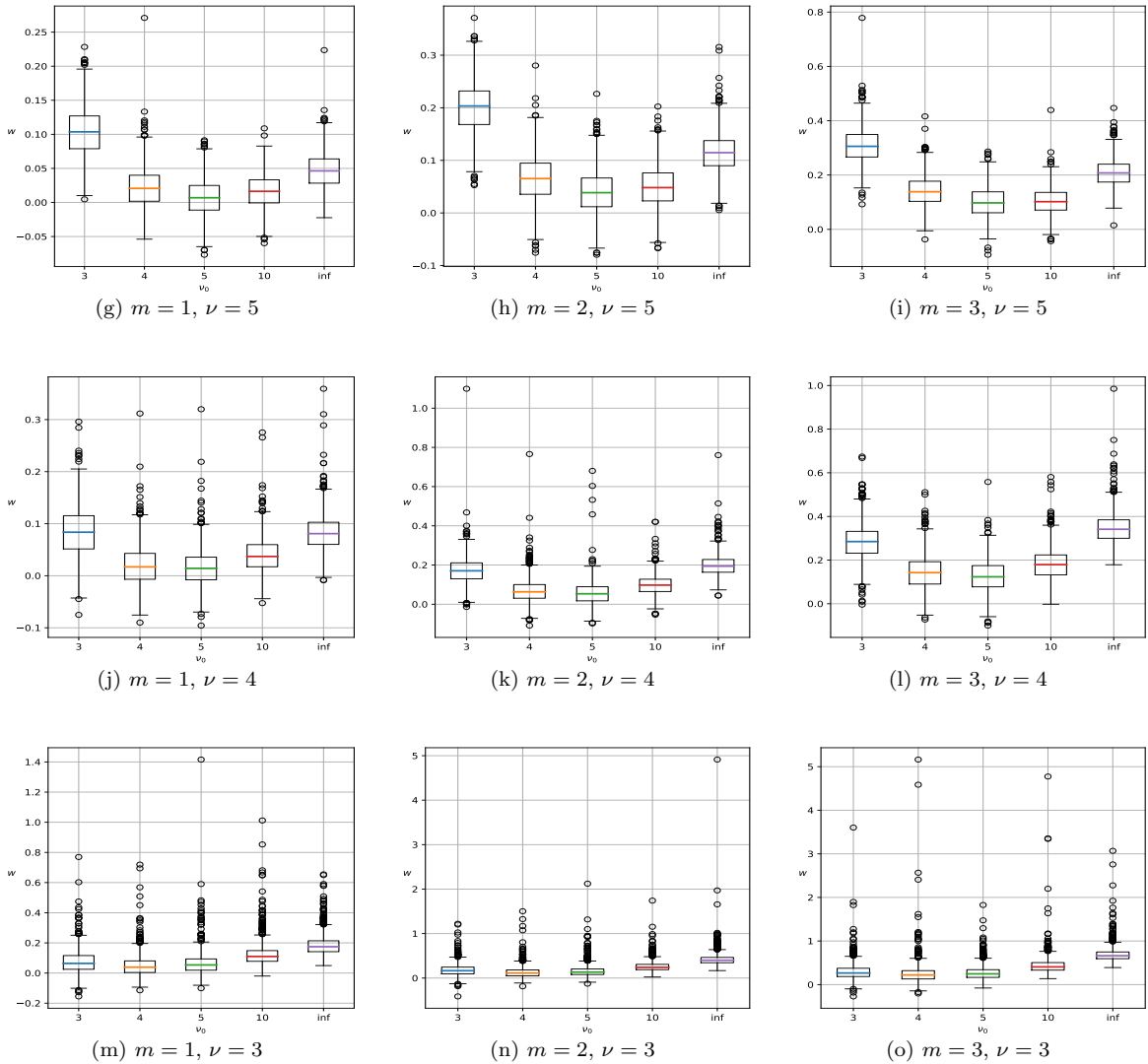

(g) $m=1, \nu=5$  (h) $m=2, \nu=5$  (i) $m=3, \nu=5$

(j) $m=1, \nu=4$  (k) $m=2, \nu=4$  (l) $m=3, \nu=4$

(m) $m=1, \nu=3$  (n) $m=2, \nu=3$  (o) $m=3, \nu=3$

### 5.2.4 Rates of convergence

We investigate the rate of convergence of the test statistic $W_{N,k}(m, \nu_0)$ on samples from $T_m(\nu)$ as the sample size $N \to \infty$ by considering the relation $W_{N,k}(m, \nu) = aN^b$ and performing linear regression on the model

$$\log W_{N,k}(m, \nu) = \log a + b \log N.$$

For every pair $\nu, \nu_0 \in \{3, 4, \ldots, 20, \infty\}$ and $N \in \{100, 200, \ldots, 5000\}$ we generate a random sample of size $N$ from the $T_m(\nu)$ distribution and compute the empirical value of $W_{N,k}(m, \nu_0)$ for $m \in \{1, 2, 3\}$ and $k = 3$. In each case, the process is repeated $M = 1000$ times, which yields a sample realisation $\{w_1, w_2, \ldots, w_M\}$ from the distribution of $W_{N,k}(m, \nu_0)$, from which we compute the sample mean $\bar{w}_{N,k}(m, \nu_0)$.

We then perform simple linear regression on the pairs $(\log N, \log \bar{w}_N)$ for $N = 100, 200, \ldots, 5000$ and record the value of the gradient as an estimate of the rate at which $W_{N,k}(m, \nu_0)$ converges to its limiting value as $N \to \infty$.

Figure 5 shows plots of $\log W_{N,k}(m, \nu_0)$ against $\log N$ on data from the $T_m(\nu)$ distribution for $\nu, \nu_0 \in \{3, 4, 5, 10, \infty\}$, $m \in \{1, 2, 3\}$ and $k = 3$. Table 1 then shows the least-squares estimates of the gradients, which in turn provide estimates for the rates of convergence.

Figure 5: $\log W_{N,k}(m, \nu_0)$ plotted against $\log N$ on samples from $T_m(\nu)$ with $k = 3$.

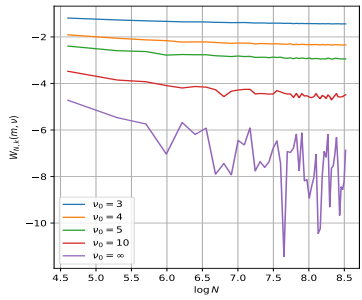
(a) $m=1, \nu=\infty$

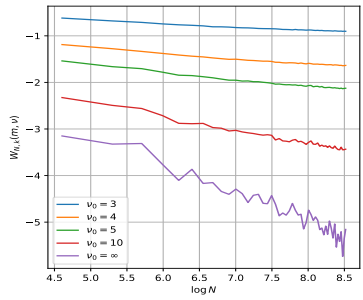
(b) $m=2, \nu=\infty$

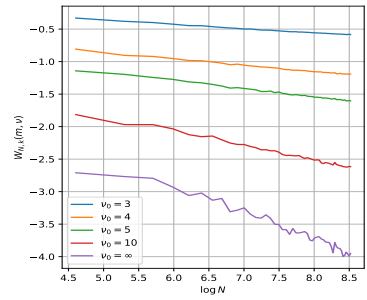
(c) $m=3, \nu=\infty$

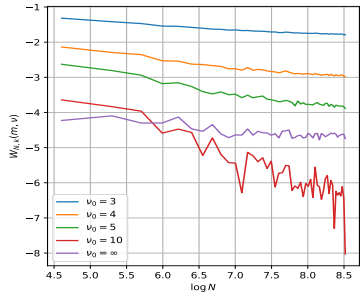
(d) $m=1, \nu=10$

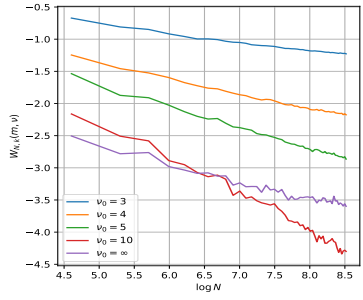
(e) $m=2, \nu=10$

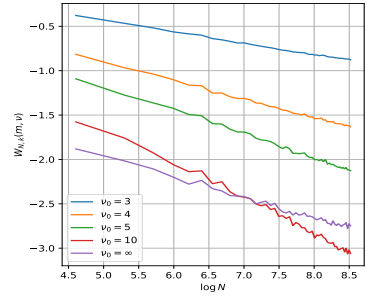
(f) $m=3, \nu=10$

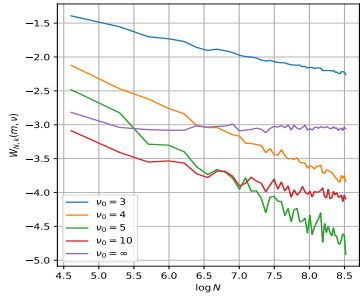
(g) $m=1, \nu=5$

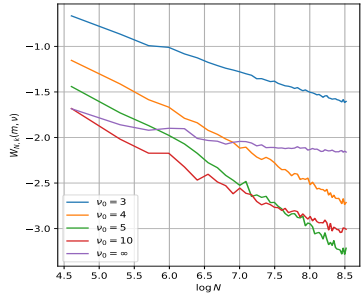
(h) $m=2, \nu=5$

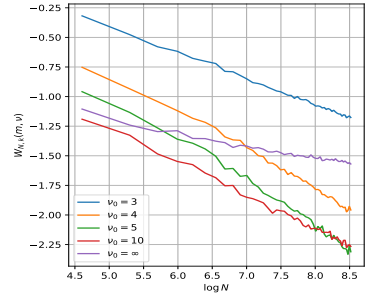
(i) $m=3, \nu=5$

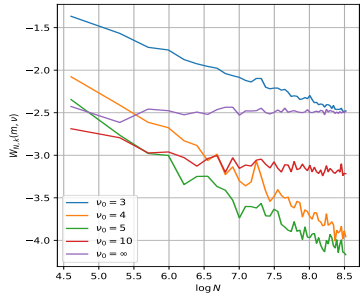
(j) $m=1, \nu=4$

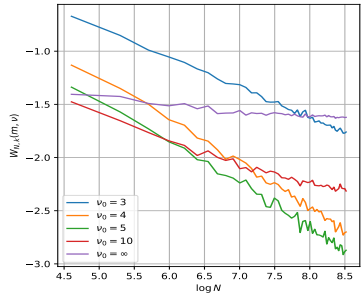
(k) $m=2, \nu=4$

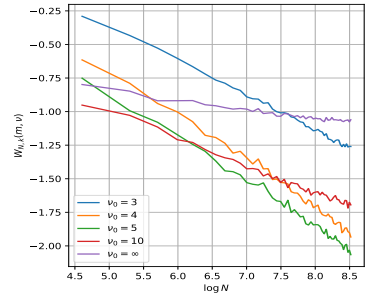
(l) $m=3, \nu=4$

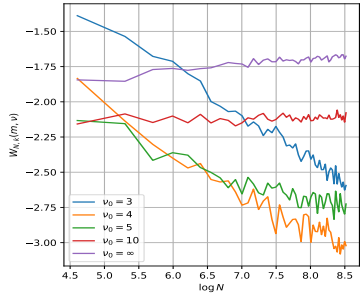
(m) $m=1, \nu=3$

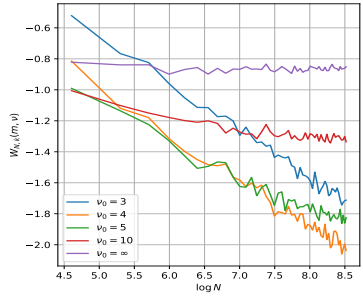
(n) $m=2, \nu=3$

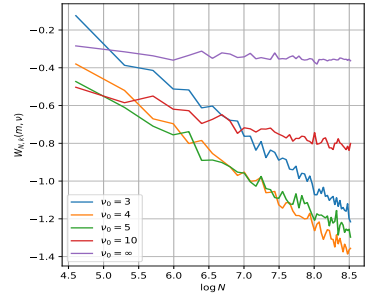
(o) $m=3, \nu=3$

Table 1: Rates of convergence of $W_{N,k}(m, \nu_0)$ on samples from $T_m(\nu)$ for $k = 3$

| $m$ | $\nu \backslash \nu_0$ | 3 | 4 | 5 | 6 | 8 | 10 | 12 | 15 | 20 | $\infty$ |
|---|---|---|---|---|---|---|---|---|---|---|---|
| 1 | 3 | -0.32 | -0.27 | -0.15 | -0.08 | -0.02 | 0.01 | 0.01 | 0.02 | 0.02 | 0.04 |
| | 4 | -0.28 | -0.48 | -0.43 | -0.31 | -0.16 | -0.10 | -0.08 | -0.06 | -0.04 | -0.00 |
| | 5 | -0.21 | -0.42 | -0.56 | -0.52 | -0.32 | -0.21 | -0.16 | -0.10 | -0.08 | -0.02 |
| | 6 | -0.17 | -0.34 | -0.52 | -0.62 | -0.53 | -0.35 | -0.27 | -0.20 | -0.14 | -0.05 |
| | 8 | -0.13 | -0.24 | -0.38 | -0.52 | -0.75 | -0.66 | -0.54 | -0.41 | -0.28 | -0.07 |
| | 10 | -0.11 | -0.20 | -0.30 | -0.40 | -0.63 | -0.80 | -0.78 | -0.61 | -0.45 | -0.13 |
| | 12 | -0.10 | -0.17 | -0.26 | -0.35 | -0.56 | -0.72 | -0.79 | -0.77 | -0.68 | -0.22 |
| | 15 | -0.09 | -0.15 | -0.21 | -0.29 | -0.42 | -0.61 | -0.71 | -0.90 | -0.74 | -0.24 |
| | 20 | -0.08 | -0.13 | -0.18 | -0.23 | -0.35 | -0.47 | -0.58 | -0.69 | -0.77 | -0.32 |
| | $\infty$ | -0.05 | -0.09 | -0.11 | -0.14 | -0.19 | -0.23 | -0.28 | -0.36 | -0.39 | -0.79 |
| 2 | 3 | -0.31 | -0.29 | -0.21 | -0.14 | -0.10 | -0.06 | -0.05 | -0.04 | -0.02 | -0.00 |
| | 4 | -0.28 | -0.42 | -0.41 | -0.34 | -0.26 | -0.20 | -0.16 | -0.13 | -0.10 | -0.05 |
| | 5 | -0.23 | -0.40 | -0.48 | -0.48 | -0.39 | -0.32 | -0.27 | -0.23 | -0.19 | -0.10 |
| | 6 | -0.19 | -0.34 | -0.46 | -0.52 | -0.49 | -0.42 | -0.36 | -0.30 | -0.25 | -0.12 |
| | 8 | -0.15 | -0.27 | -0.38 | -0.47 | -0.55 | -0.55 | -0.51 | -0.43 | -0.37 | -0.20 |
| | 10 | -0.13 | -0.23 | -0.32 | -0.41 | -0.54 | -0.57 | -0.57 | -0.54 | -0.48 | -0.26 |
| | 12 | -0.12 | -0.21 | -0.29 | -0.37 | -0.49 | -0.56 | -0.60 | -0.59 | -0.56 | -0.33 |
| | 15 | -0.11 | -0.18 | -0.26 | -0.32 | -0.44 | -0.52 | -0.57 | -0.60 | -0.59 | -0.38 |
| | 20 | -0.10 | -0.16 | -0.22 | -0.28 | -0.38 | -0.47 | -0.53 | -0.57 | -0.62 | -0.46 |
| | $\infty$ | -0.07 | -0.11 | -0.14 | -0.18 | -0.24 | -0.28 | -0.33 | -0.38 | -0.43 | -0.61 |
| 3 | 3 | -0.27 | -0.26 | -0.20 | -0.16 | -0.12 | -0.08 | -0.07 | -0.05 | -0.04 | -0.01 |
| | 4 | -0.26 | -0.35 | -0.33 | -0.31 | -0.24 | -0.20 | -0.18 | -0.15 | -0.12 | -0.07 |
| | 5 | -0.22 | -0.33 | -0.37 | -0.37 | -0.33 | -0.28 | -0.25 | -0.22 | -0.18 | -0.10 |
| | 6 | -0.19 | -0.29 | -0.36 | -0.39 | -0.38 | -0.35 | -0.31 | -0.27 | -0.24 | -0.13 |
| | 8 | -0.15 | -0.24 | -0.32 | -0.36 | -0.40 | -0.40 | -0.38 | -0.36 | -0.32 | -0.19 |
| | 10 | -0.13 | -0.21 | -0.27 | -0.32 | -0.38 | -0.40 | -0.40 | -0.40 | -0.36 | -0.23 |
| | 12 | -0.12 | -0.19 | -0.24 | -0.29 | -0.36 | -0.39 | -0.40 | -0.40 | -0.38 | -0.26 |
| | 15 | -0.10 | -0.17 | -0.22 | -0.26 | -0.33 | -0.36 | -0.39 | -0.41 | -0.41 | -0.30 |
| | 20 | -0.09 | -0.15 | -0.20 | -0.24 | -0.30 | -0.34 | -0.37 | -0.39 | -0.40 | -0.34 |
| | $\infty$ | -0.06 | -0.10 | -0.13 | -0.15 | -0.19 | -0.22 | -0.25 | -0.27 | -0.31 | -0.38 |

When $\nu$ and $\nu_0$ are both moderately large, Figure 5 amd Table 1 indicate that rates of convergence decrease as the dimension $m$ increases. By contrast, when $\nu$ and $\nu_0$ are both small, rates of convergence are also small, and do not appear to depend on the dimsnsion.

Figure 5 and Table 1 reveal that rates of convergence are greater under the null hypothesis $\nu = \nu_0$ (corresponding to the diagonal entries of Table 1) compared with the alternative $\nu \neq \nu_0$, and moreover that the rate of convergence decreases as the difference between $\nu$ and $\nu_0$ increases. This is perhaps because, when one of $\nu$ and $\nu_0$ is small and the other is large, the test statistic converges quickly to a limiting value and remain stable thereafter, as shown by the first and last rows of Figure 2.

### 5.2.5 Critical values

For significance level $\alpha$ we estimate the upper-tail critical value $w_\alpha$ of the $W_{N,k}(m, \nu)$ distribution, defined to be the value for which $\mathbb{P}(W_{N,k} \geq w_\alpha) = \alpha$, by the empirical quantile $\hat{w}_\alpha = w_{(N[1-\alpha])}$ where $w_{(j)}$ is the $j$th order statistic (or a suitable combination of adjacent order statistics) of the associated sample realisation $\{w_1, w_2, \ldots, w_M\}$.

Table 2 shows estimated critical values $\hat{w}_\alpha$ for the distribution of $W_{N,k}(m, \nu)$ at significance level $\alpha = 0.05$ for $m \in \{1, 2, 3\}$, $k \in \{1, 2, 3\}$ and $\nu \in \{3, 5, 10, \infty\}$, for sample sizes of $N \in \{100, 200, \ldots, 900, 1000, 2000, \ldots, 5000\}$. The results indicate that critical values of $W_{N,k}(m, \nu)$ increase as the dimension $m$ increases, and decrease as the shape parameter $\nu$ and sample size $N$ increase, all of which can be verified by inspection of the empirical densities shown in Figures 3 and 4.

Table 2: Estimated critical values of $W_{N,k}(m,\nu)$ at $\alpha = 0.05$.

| | | $m=1$ | | | $m=2$ | | | $m=3$ | | |
|---|---|---|---|---|---|---|---|---|---|---|
| $\nu$ | $N$ | $k=1$ | $k=2$ | $k=3$ | $k=1$ | $k=2$ | $k=3$ | $k=1$ | $k=2$ | $k=3$ |
| 3 | 100 | 0.640 | 0.635 | 0.660 | 1.217 | 1.307 | 1.349 | 1.853 | 1.916 | 2.025 |
| | 200 | 0.562 | 0.553 | 0.590 | 0.960 | 1.005 | 1.075 | 1.330 | 1.423 | 1.508 |
| | 300 | 0.469 | 0.471 | 0.507 | 0.880 | 0.930 | 0.979 | 1.288 | 1.375 | 1.415 |
| | 400 | 0.452 | 0.435 | 0.465 | 0.772 | 0.823 | 0.859 | 1.139 | 1.209 | 1.266 |
| | 500 | 0.413 | 0.421 | 0.422 | 0.721 | 0.747 | 0.800 | 1.139 | 1.216 | 1.254 |
| | 600 | 0.392 | 0.408 | 0.420 | 0.681 | 0.720 | 0.765 | 0.970 | 1.059 | 1.114 |
| | 700 | 0.363 | 0.374 | 0.390 | 0.670 | 0.721 | 0.751 | 1.069 | 1.155 | 1.198 |
| | 800 | 0.347 | 0.346 | 0.361 | 0.629 | 0.683 | 0.722 | 0.960 | 1.007 | 1.064 |
| | 900 | 0.336 | 0.330 | 0.340 | 0.629 | 0.668 | 0.686 | 0.924 | 0.994 | 1.036 |
| | 1000 | 0.337 | 0.340 | 0.357 | 0.582 | 0.609 | 0.622 | 0.919 | 0.961 | 1.018 |
| | 2000 | 0.270 | 0.286 | 0.298 | 0.467 | 0.500 | 0.518 | 0.703 | 0.769 | 0.808 |
| | 3000 | 0.245 | 0.240 | 0.249 | 0.409 | 0.435 | 0.463 | 0.622 | 0.661 | 0.704 |
| | 4000 | 0.216 | 0.230 | 0.228 | 0.385 | 0.408 | 0.428 | 0.582 | 0.630 | 0.670 |
| | 5000 | 0.191 | 0.199 | 0.199 | 0.368 | 0.388 | 0.407 | 0.526 | 0.573 | 0.605 |
| 5 | 100 | 0.373 | 0.369 | 0.371 | 0.621 | 0.611 | 0.655 | 0.862 | 0.870 | 0.896 |
| | 200 | 0.266 | 0.273 | 0.268 | 0.444 | 0.455 | 0.488 | 0.658 | 0.703 | 0.734 |
| | 300 | 0.212 | 0.209 | 0.209 | 0.370 | 0.395 | 0.412 | 0.558 | 0.593 | 0.629 |
| | 400 | 0.199 | 0.191 | 0.189 | 0.334 | 0.351 | 0.381 | 0.456 | 0.490 | 0.529 |
| | 500 | 0.164 | 0.168 | 0.167 | 0.294 | 0.328 | 0.332 | 0.457 | 0.500 | 0.515 |
| | 600 | 0.151 | 0.148 | 0.152 | 0.259 | 0.272 | 0.291 | 0.416 | 0.464 | 0.493 |
| | 700 | 0.148 | 0.144 | 0.144 | 0.260 | 0.264 | 0.271 | 0.379 | 0.420 | 0.454 |
| | 800 | 0.139 | 0.138 | 0.145 | 0.234 | 0.238 | 0.254 | 0.375 | 0.389 | 0.419 |
| | 900 | 0.134 | 0.131 | 0.129 | 0.221 | 0.234 | 0.244 | 0.343 | 0.370 | 0.394 |
| | 1000 | 0.128 | 0.128 | 0.131 | 0.211 | 0.219 | 0.232 | 0.328 | 0.355 | 0.384 |
| | 2000 | 0.093 | 0.087 | 0.086 | 0.148 | 0.158 | 0.168 | 0.243 | 0.269 | 0.291 |
| | 3000 | 0.077 | 0.075 | 0.073 | 0.126 | 0.133 | 0.141 | 0.193 | 0.216 | 0.235 |
| | 4000 | 0.064 | 0.063 | 0.060 | 0.106 | 0.112 | 0.118 | 0.179 | 0.202 | 0.214 |
| | 5000 | 0.058 | 0.058 | 0.058 | 0.103 | 0.106 | 0.111 | 0.158 | 0.175 | 0.190 |
| 10 | 100 | 0.298 | 0.268 | 0.256 | 0.416 | 0.428 | 0.425 | 0.589 | 0.592 | 0.585 |
| | 200 | 0.210 | 0.198 | 0.196 | 0.300 | 0.304 | 0.303 | 0.407 | 0.439 | 0.452 |
| | 300 | 0.166 | 0.158 | 0.152 | 0.255 | 0.251 | 0.259 | 0.325 | 0.345 | 0.370 |
| | 400 | 0.136 | 0.125 | 0.126 | 0.206 | 0.211 | 0.216 | 0.288 | 0.303 | 0.322 |
| | 500 | 0.140 | 0.125 | 0.119 | 0.197 | 0.194 | 0.192 | 0.261 | 0.272 | 0.294 |
| | 600 | 0.125 | 0.103 | 0.104 | 0.172 | 0.166 | 0.173 | 0.246 | 0.260 | 0.269 |
| | 700 | 0.112 | 0.104 | 0.096 | 0.166 | 0.166 | 0.174 | 0.229 | 0.246 | 0.259 |
| | 800 | 0.112 | 0.089 | 0.092 | 0.145 | 0.140 | 0.152 | 0.223 | 0.235 | 0.251 |
| | 900 | 0.098 | 0.090 | 0.087 | 0.146 | 0.144 | 0.149 | 0.197 | 0.210 | 0.217 |
| | 1000 | 0.087 | 0.082 | 0.078 | 0.131 | 0.131 | 0.132 | 0.198 | 0.206 | 0.220 |
| | 2000 | 0.065 | 0.057 | 0.055 | 0.098 | 0.096 | 0.097 | 0.139 | 0.147 | 0.157 |
| | 3000 | 0.057 | 0.048 | 0.043 | 0.075 | 0.072 | 0.078 | 0.106 | 0.119 | 0.128 |
| | 4000 | 0.045 | 0.039 | 0.037 | 0.066 | 0.066 | 0.067 | 0.103 | 0.106 | 0.115 |
| | 5000 | 0.041 | 0.035 | 0.034 | 0.058 | 0.057 | 0.058 | 0.092 | 0.101 | 0.107 |
| $\infty$ | 100 | 0.282 | 0.223 | 0.223 | 0.360 | 0.306 | 0.314 | 0.426 | 0.378 | 0.368 |
| | 200 | 0.205 | 0.160 | 0.156 | 0.275 | 0.236 | 0.226 | 0.302 | 0.295 | 0.281 |
| | 300 | 0.171 | 0.132 | 0.121 | 0.213 | 0.193 | 0.188 | 0.262 | 0.256 | 0.245 |
| | 400 | 0.142 | 0.114 | 0.107 | 0.187 | 0.162 | 0.160 | 0.232 | 0.215 | 0.217 |
| | 500 | 0.127 | 0.107 | 0.097 | 0.161 | 0.137 | 0.134 | 0.192 | 0.190 | 0.187 |
| | 600 | 0.124 | 0.100 | 0.090 | 0.161 | 0.140 | 0.135 | 0.186 | 0.172 | 0.176 |
| | 700 | 0.119 | 0.091 | 0.085 | 0.135 | 0.117 | 0.112 | 0.169 | 0.164 | 0.166 |
| | 800 | 0.102 | 0.085 | 0.074 | 0.124 | 0.111 | 0.108 | 0.155 | 0.147 | 0.147 |
| | 900 | 0.098 | 0.079 | 0.071 | 0.121 | 0.103 | 0.102 | 0.152 | 0.150 | 0.147 |
| | 1000 | 0.094 | 0.071 | 0.064 | 0.119 | 0.100 | 0.108 | 0.144 | 0.133 | 0.138 |
| | 2000 | 0.068 | 0.048 | 0.045 | 0.082 | 0.071 | 0.069 | 0.105 | 0.097 | 0.097 |
| | 3000 | 0.056 | 0.041 | 0.039 | 0.067 | 0.057 | 0.056 | 0.081 | 0.078 | 0.080 |
| | 4000 | 0.046 | 0.037 | 0.034 | 0.061 | 0.054 | 0.050 | 0.074 | 0.072 | 0.071 |
| | 5000 | 0.043 | 0.033 | 0.030 | 0.049 | 0.044 | 0.041 | 0.065 | 0.062 | 0.064 |

### 5.2.6 Statistical power

For the null hypothesis $\nu = \nu_0$, the power of the test to detect the alternative hypothesis $\nu \neq \nu_0$ at significance level $\alpha$ is defined by
$$\gamma(\nu) = \mathbb{P}_\nu\big(W_{N,k}(m,\nu_0) > w_\alpha\big), \qquad \nu \neq \nu_0,$$
where $w_\alpha$ is the $(1-\alpha)$-quantile of the distribution of $W_{N,k}(m,\nu_0)$ under the null hypothesis, and $\mathbb{P}_\nu$ is defined by distribution of $W_{N,k}(m,\nu_0)$ on samples from the $T_m(\nu)$ distribution.

We estimate $\gamma(\nu)$ using the estimated critical values $\hat{w}_\alpha$ computed in the previous section: let $\{w_1, w_2, \ldots, w_M\}$ independent observations of the test statistic $W_{N,k}(m,\nu_0)$ on samples from $T_m(\nu)$, and consider the estimator
$$\hat{\gamma}(\nu) = \frac{1}{M}\sum_{j=1}^M I(w_j > \hat{w}_\alpha)$$
where $I$ is the indicator function. This is the proportion of observations that exceed the estimated critical value $\hat{w}_\alpha$, and serves as an estimate for the probability that the test correctly rejects the null hypothesis $\nu = \nu_0$ in favour of the alternative $\nu \neq \nu_0$. Table 3 shows the estimated power for $N = 5000$ and $k = 3$ at significance level $\alpha = 0.05$.

Table 3: Statistical power of $W_{N,k}(m,\nu_0)$ on samples from $T_m(\nu)$ for $N = 5000, k = 3, \alpha = 0.05$

| $m$ | $\nu_0$ \ $\nu$ | 3 | 4 | 5 | 6 | 8 | 10 | 12 | 15 | 20 | $\infty$ |
|---|---|---|---|---|---|---|---|---|---|---|---|
| 1 | 3 | 0.05 | 0.03 | 0.04 | 0.04 | 0.08 | 0.10 | 0.11 | 0.14 | 0.15 | 0.33 |
|   | 4 | 0.50 | 0.05 | 0.04 | 0.04 | 0.07 | 0.10 | 0.10 | 0.17 | 0.24 | 0.48 |
|   | 5 | 0.90 | 0.10 | 0.05 | 0.03 | 0.05 | 0.06 | 0.08 | 0.11 | 0.13 | 0.34 |
|   | 6 | 1.00 | 0.27 | 0.07 | 0.05 | 0.04 | 0.05 | 0.07 | 0.09 | 0.11 | 0.27 |
|   | 8 | 1.00 | 0.56 | 0.15 | 0.07 | 0.05 | 0.06 | 0.05 | 0.04 | 0.06 | 0.12 |
|   | 10 | 1.00 | 0.78 | 0.25 | 0.10 | 0.06 | 0.05 | 0.05 | 0.05 | 0.06 | 0.10 |
|   | 12 | 1.00 | 0.87 | 0.36 | 0.17 | 0.08 | 0.06 | 0.05 | 0.05 | 0.05 | 0.10 |
|   | 15 | 1.00 | 0.94 | 0.48 | 0.21 | 0.09 | 0.07 | 0.06 | 0.05 | 0.06 | 0.09 |
|   | 20 | 1.00 | 0.98 | 0.58 | 0.27 | 0.10 | 0.04 | 0.05 | 0.04 | 0.05 | 0.06 |
|   | $\infty$ | 1.00 | 1.00 | 0.89 | 0.59 | 0.23 | 0.15 | 0.10 | 0.08 | 0.06 | 0.05 |
| 2 | 3 | 0.05 | 0.03 | 0.04 | 0.03 | 0.07 | 0.09 | 0.12 | 0.14 | 0.20 | 0.46 |
|   | 4 | 0.51 | 0.05 | 0.02 | 0.04 | 0.06 | 0.08 | 0.12 | 0.20 | 0.29 | 0.71 |
|   | 5 | 0.97 | 0.16 | 0.05 | 0.03 | 0.04 | 0.06 | 0.09 | 0.11 | 0.17 | 0.55 |
|   | 6 | 1.00 | 0.40 | 0.10 | 0.05 | 0.04 | 0.05 | 0.06 | 0.07 | 0.12 | 0.39 |
|   | 8 | 1.00 | 0.86 | 0.31 | 0.12 | 0.05 | 0.06 | 0.04 | 0.06 | 0.08 | 0.22 |
|   | 10 | 1.00 | 0.96 | 0.49 | 0.21 | 0.07 | 0.05 | 0.05 | 0.05 | 0.06 | 0.14 |
|   | 12 | 1.00 | 0.99 | 0.68 | 0.31 | 0.11 | 0.07 | 0.05 | 0.06 | 0.05 | 0.10 |
|   | 15 | 1.00 | 1.00 | 0.81 | 0.44 | 0.15 | 0.10 | 0.07 | 0.05 | 0.06 | 0.08 |
|   | 20 | 1.00 | 1.00 | 0.93 | 0.58 | 0.18 | 0.10 | 0.07 | 0.05 | 0.05 | 0.05 |
|   | $\infty$ | 1.00 | 1.00 | 1.00 | 0.97 | 0.60 | 0.35 | 0.23 | 0.16 | 0.13 | 0.05 |
| 3 | 3 | 0.05 | 0.03 | 0.04 | 0.05 | 0.08 | 0.12 | 0.15 | 0.19 | 0.27 | 0.72 |
|   | 4 | 0.51 | 0.05 | 0.02 | 0.04 | 0.05 | 0.08 | 0.12 | 0.17 | 0.32 | 0.84 |
|   | 5 | 0.98 | 0.18 | 0.05 | 0.03 | 0.04 | 0.03 | 0.08 | 0.09 | 0.19 | 0.62 |
|   | 6 | 1.00 | 0.47 | 0.11 | 0.05 | 0.03 | 0.03 | 0.04 | 0.06 | 0.11 | 0.45 |
|   | 8 | 1.00 | 0.91 | 0.36 | 0.14 | 0.05 | 0.04 | 0.03 | 0.03 | 0.05 | 0.20 |
|   | 10 | 1.00 | 0.99 | 0.64 | 0.28 | 0.07 | 0.05 | 0.04 | 0.03 | 0.04 | 0.10 |
|   | 12 | 1.00 | 1.00 | 0.86 | 0.47 | 0.14 | 0.07 | 0.05 | 0.04 | 0.03 | 0.08 |
|   | 15 | 1.00 | 1.00 | 0.96 | 0.68 | 0.21 | 0.10 | 0.07 | 0.05 | 0.04 | 0.06 |
|   | 20 | 1.00 | 1.00 | 0.99 | 0.82 | 0.33 | 0.17 | 0.09 | 0.08 | 0.05 | 0.06 |
|   | $\infty$ | 1.00 | 1.00 | 1.00 | 1.00 | 0.89 | 0.65 | 0.40 | 0.26 | 0.15 | 0.05 |

Table 3 indicates that the power to detect $\nu \neq \nu_0$ increases as $|\nu - \nu_0|$ increases. We note an asymmetry either side of the main diagonal: the test has good power to detect $\nu \neq \nu_0$ when $\nu_0$ is small, but less so when $\nu_0$ is large. To account for this, Figure 4 shows that samples from the distribution of $W_{N,k}(m,\nu_0)$ contain an increasing number of outlying observations as $\nu$ decreases and the associated distribution $T_m(\nu)$ becomes increasingly heavy-tailed. This leads to an increase in the variance of $W_{N,k}(m,\nu_0)$ and a corresponding decrease in the power of the test to detect $\nu \neq \nu_0$ as $\nu$ decreases. By contrast, when $\nu$ is large, for example when samples are drawn from the multivariate Gaussian distribution $T_m(\infty)$, there are relatively few outlying values of $W_{N,k}(m,\nu_0)$, and the resulting separation between the empirical distributions indicates that power of the test to detect large values of $\nu$ when $\nu_0$ is small, is relatively good.

## 5.3 Numerical experiments for $W^*_{N,k}(m,\eta)$

We now repeat numerical experiments for the Pearson II test statistic $W^*_{N,k}(m,\eta)$, defined in (21). Let $X \sim P_m(\eta)$ where $\eta > 0$ is unknown. For $\eta_0 > 0$ fixed, we test the null hypothesis $\eta = \eta_0$ against the alternative hypothesis $\eta \neq \eta_0$, based on the value of $W^*_{N,k}(m,\eta_0)$. By (23), large values indicate that the null hypothesis should be rejected.

**Summary of results**

- $W^*_{N,k}(m,\eta_0)$ on samples from $P_m(\eta)$ appears to onverge to zero when $\eta = \eta_0$, and to a strictly positive constant when $\eta \neq \eta_0$, as the sample size $N \to \infty$, which verifies (23).

- The rate at which $W^*_{N,k}(m,\eta_0)$ converges appears to decreases with the dimension $m$. Under the null hypothesis $\eta = \eta_0$, the rate appears to increase as $\eta$ increases. Under the alternative hypothesis $\eta \neq \eta_0$, the rate appears to decrease as the difference $|\eta - \eta_0|$ increases.

- The statistical power of the associated hypothesis test to detect $\eta \neq \eta_0$ appears to increase as the dimension $m$ increases, and also as the difference $|\eta - \eta_0|$ increases. Overall the power of the test is low compared with that of the test based on $W_{N,k}(m,\nu_0)$, which is perhaps because the sample variance of $W^*_{N,k}(m,\eta_0)$ is relatively large compared to the differences between its limiting values for different values of $\eta \neq \eta_0$

### 5.3.1 Consistency

Figure 6 shows the asymptotic behaviour of the test statistic $W^*_{N,k}(m,\eta)$, for $m \in \{1,2,3\}$ and $k \in \{1,2,3,4\}$, on samples drawn from the $P_m(\eta)$ distribution with $\eta \in \{3,4,5,10,\infty\}$, as the sample size $N \to \infty$. In each plot, the lines represent the sample mean of the observed values.

Figure 6: Consistency of $W^*_{N,k}(m,\eta)$ for $k \in \{1,2,3,4\}$.

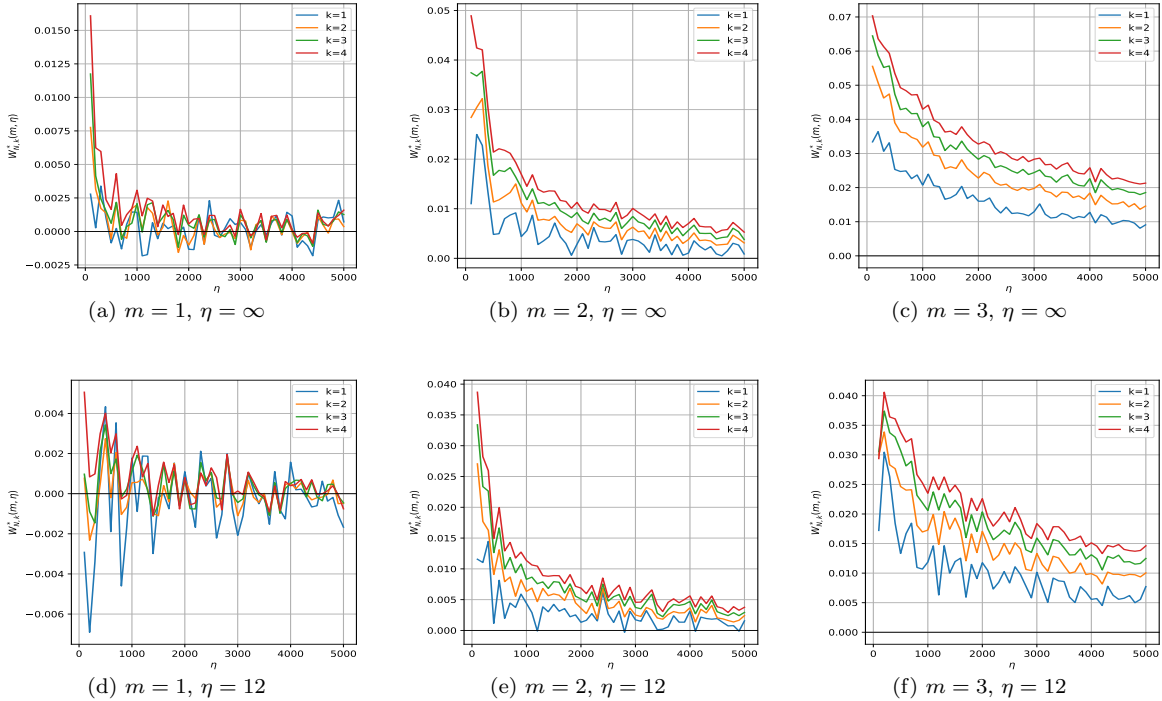

(a) $m=1$, $\eta=\infty$  (b) $m=2$, $\eta=\infty$  (c) $m=3$, $\eta=\infty$

(d) $m=1$, $\eta=12$  (e) $m=2$, $\eta=12$  (f) $m=3$, $\eta=12$

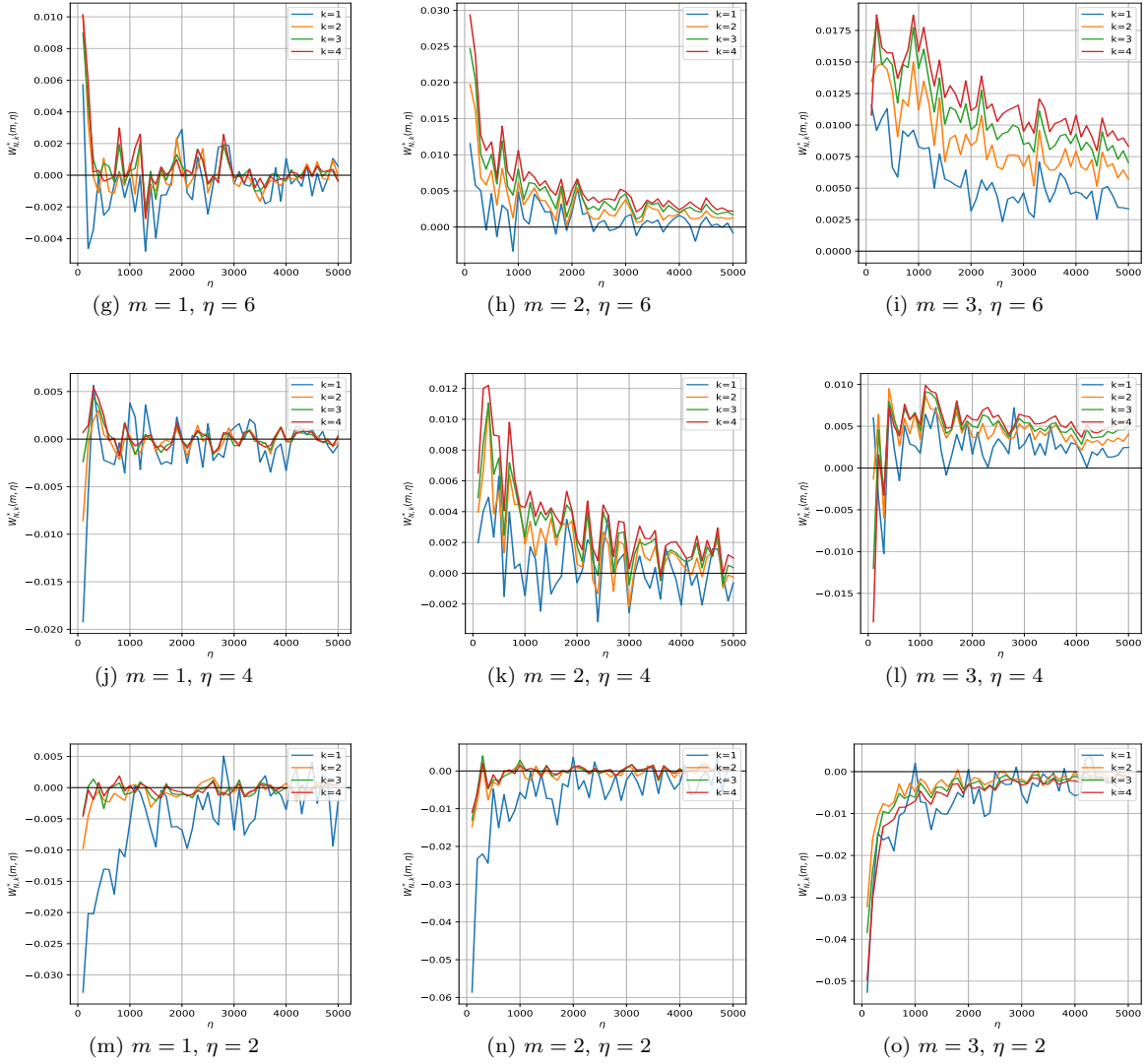

(g) $m=1, \eta=6$     (h) $m=2, \eta=6$     (i) $m=3, \eta=6$

(j) $m=1, \eta=4$     (k) $m=2, \eta=4$     (l) $m=3, \eta=4$

(m) $m=1, \eta=2$     (n) $m=2, \eta=2$     (o) $m=3, \eta=2$

As in Figure 1 for the Student statistic $W_{N,k}(m,\nu)$, Figure 6 indicates that the rate at which the Pearson II statistic $W^*_{N,k}(m,\eta)$ converges to zero increases as the parameter $\eta$ increases (or equivalently as the data become increasingly Gaussian), and decreases as the dimension $m$ increases.

Comparing Figure 6 for $W^*_{N,k}(m,\eta)$ to Figure 1 for $W_{N,k}(m,\nu)$, we see that the first row of plots are similar, which is to be expected given that both are computed on samples from the multivariate Gaussian distribution. In contrast, Figure 6 indicates that the rate at which $W^*_{N,k}(m,\eta)$ increases as $\eta$ decreases and $P_m(\eta)$ becomes increasingly light-talied.

### 5.3.2 Convergence

Figure 7 shows the asymptotic behaviour of the test statistic $W^*_{N,k}(m,\eta_0)$ for $m \in \{1,2,3\}$ and $k=3$ on samples drawn from the $P_m(\eta)$ distribution, with $\eta, \eta_0 \in \{2,4,6,10,\infty\}$, as the sample size $N \to \infty$. In each plot, the lines represent the sample mean of the observed values, with the length of the error bars equal to the sample standard deviation.

Figure 7: Convergence of $W^*_{N,k}(m,\eta)$ on samples from $P_m(\eta)$ for $k=3$.

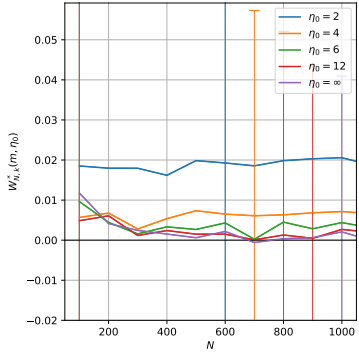
(a) $m=1$, $\eta=\infty$

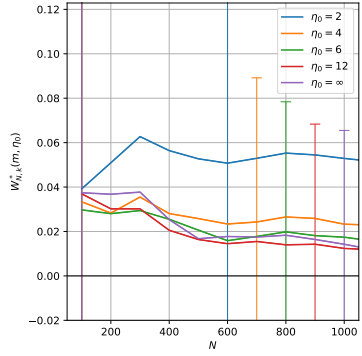
(b) $m=2$, $\eta=\infty$

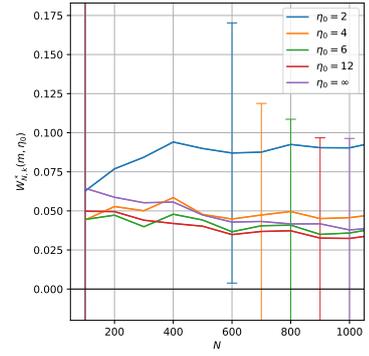
(c) $m=3$, $\eta=\infty$

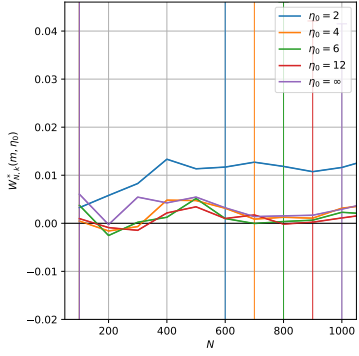
(d) $m=1$, $\eta=12$

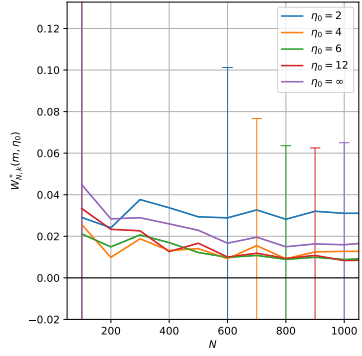
(e) $m=2$, $\eta=12$

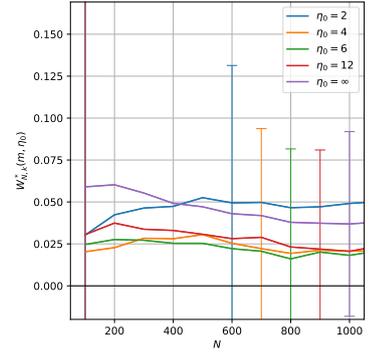
(f) $m=3$, $\eta=12$

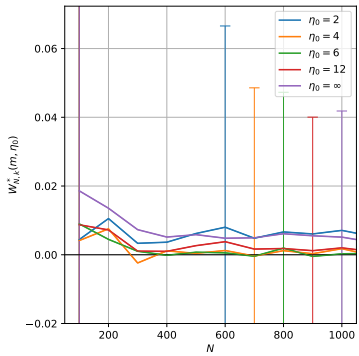
(g) $m=1$, $\eta=6$

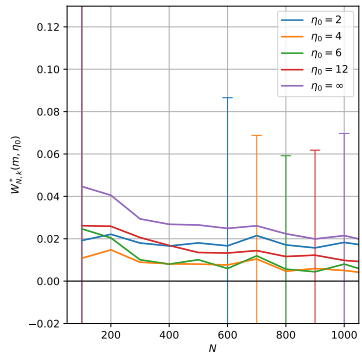
(h) $m=2$, $\eta=6$

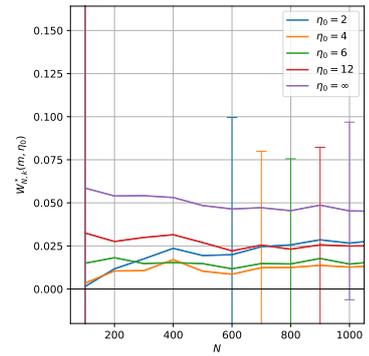
(i) $m=3$, $\eta=6$

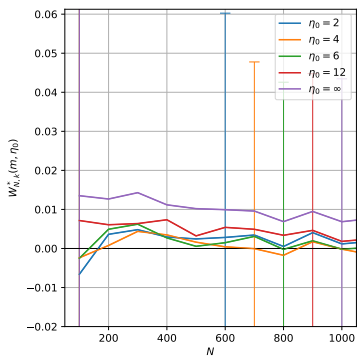
(j) $m=1$, $\eta=4$

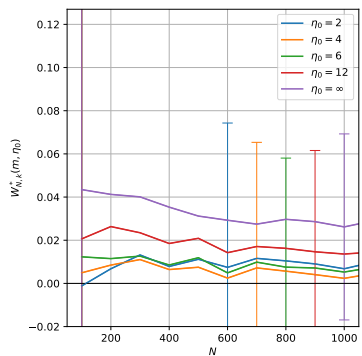
(k) $m=2$, $\eta=4$

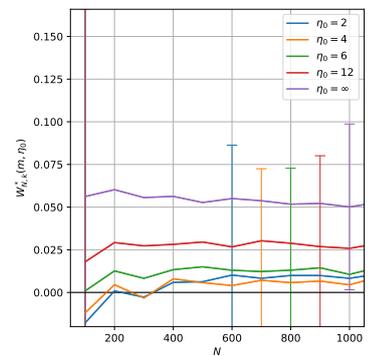
(l) $m=3$, $\eta=4$

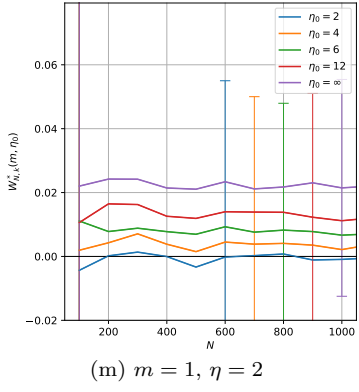
(m) $m = 1$, $\eta = 2$

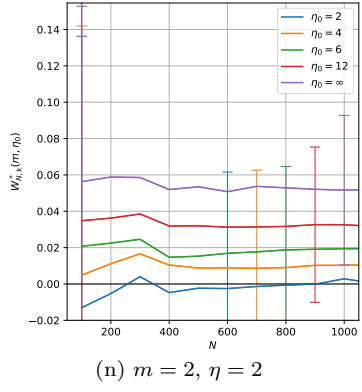
(n) $m = 2$, $\eta = 2$

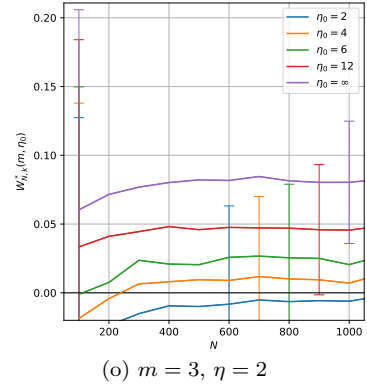
(o) $m = 3$, $\eta = 2$

In contrast to Figure 2, which shows the convergence of $W_{N,k}(m, \nu_0)$ on samples from $T_m(\nu)$, Figure 6 shows that $W^*_{N,k}(m, \eta_0)$ on samples from $P_m(\eta)$ converges rapidly, and the rate of convergence increases as $\eta$ decreases and $P_m(\eta)$ becomes increasingly light-tailed. Unfortunately, the differences between these limiting values are small compared to the variance of $W^*_{N,k}(m, \eta)$, so the power of the test to detect $\eta \neq \eta_0$ is likely to be poor.

### 5.3.3 Empirical densities

Figure 8 shows the empirical density functions of the test statistic $W^*_{N,k}(m, \eta_0)$ on data from the $P(\eta)$ distribution, for $\eta, \eta_0 \in \{3, 4, 5, 10, \infty\}$ with $m \in \{1, 2, 3\}$, $N = 5000$ and $k = 3$.

Figure 8: Empirical density of $W^*_{N,k}(m, \eta_0)$ on samples from $P_m(\eta)$ for $N = 5000$ and $k = 3$.

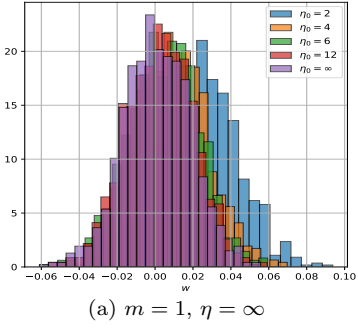
(a) $m = 1$, $\eta = \infty$

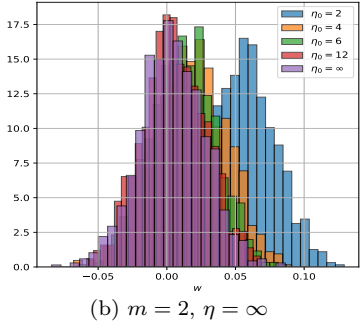
(b) $m = 2$, $\eta = \infty$

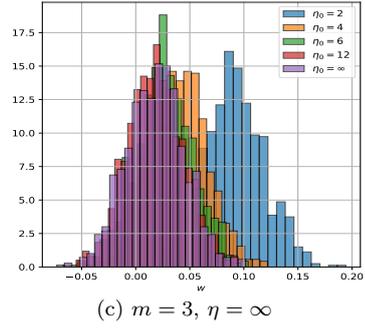
(c) $m = 3$, $\eta = \infty$

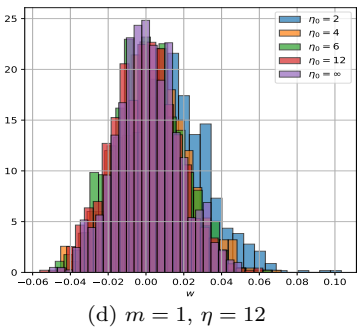
(d) $m = 1$, $\eta = 12$

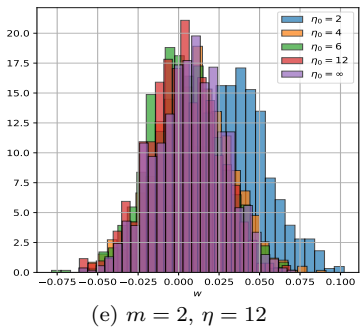
(e) $m = 2$, $\eta = 12$

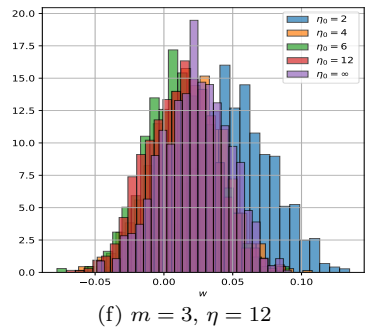
(f) $m = 3$, $\eta = 12$

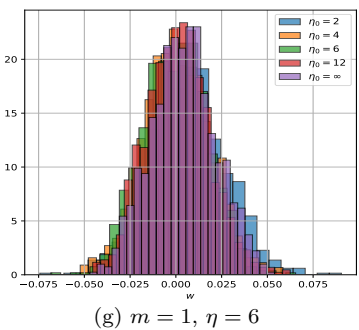
(g) $m = 1$, $\eta = 6$

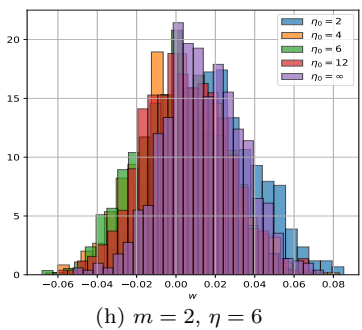
(h) $m = 2$, $\eta = 6$

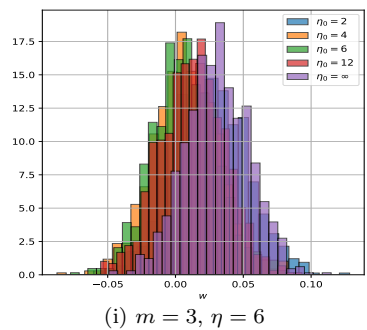
(i) $m = 3$, $\eta = 6$

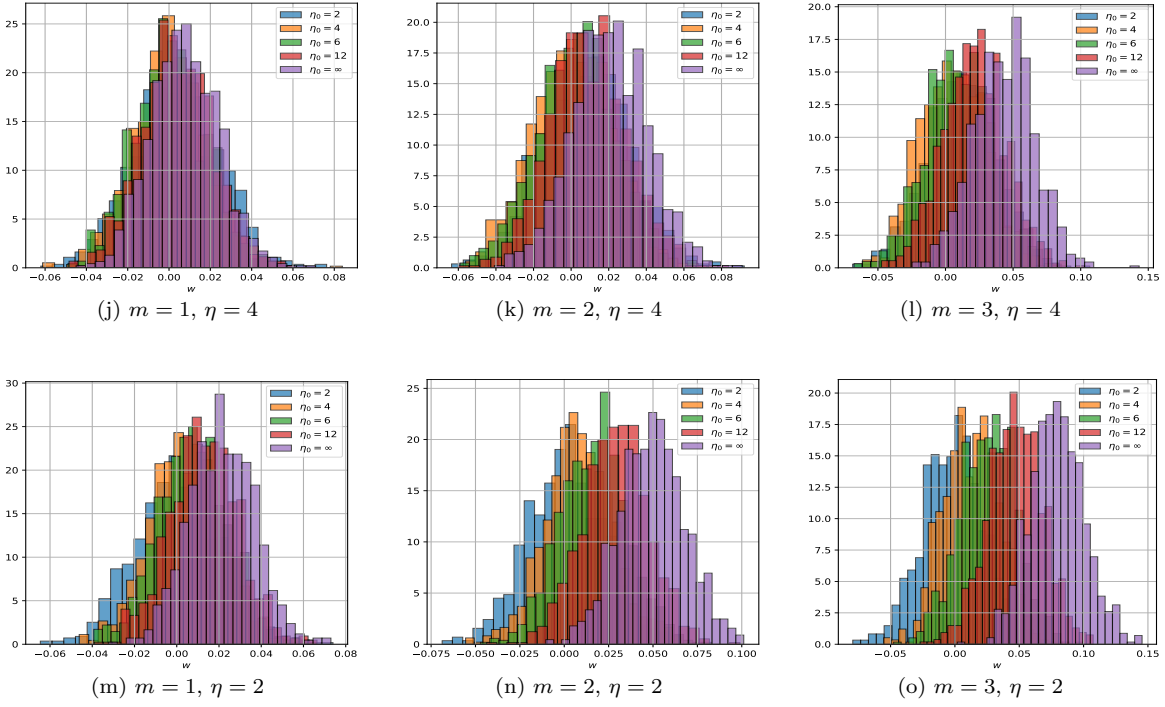

(j) $m=1$, $\eta=4$    (k) $m=2$, $\eta=4$    (l) $m=3$, $\eta=4$

(m) $m=1$, $\eta=2$    (n) $m=2$, $\eta=2$    (o) $m=3$, $\eta=2$

Compared with Figure 3, Figure 8 shows relatively little separation between the empirical densities of $W_{N,k}^*(m,\eta_0)$ on samples from $P_m(\eta)$ for any $\eta \in \{2,4,6,12,\infty\}$, indicating that power of the test to detect $\eta \neq \eta_0$ is likely to be poor. It is also evident that the separation increases as $|\eta - \eta_0|$ increases.

Figure 9 shows box plots corresponding to the empirical density functions presented in Figure 8. Compared with Figure 4, here we see little separation between the empirical means of $W_{N,k}^*(m,\eta)$ compared with their variances, even though there are no outliers in the sample values.

Figure 9: Box plots of $W_{N,k}^*(m,\eta_0)$ on samples from $P_m(\eta)$ for $N=5000$ and $k=3$.

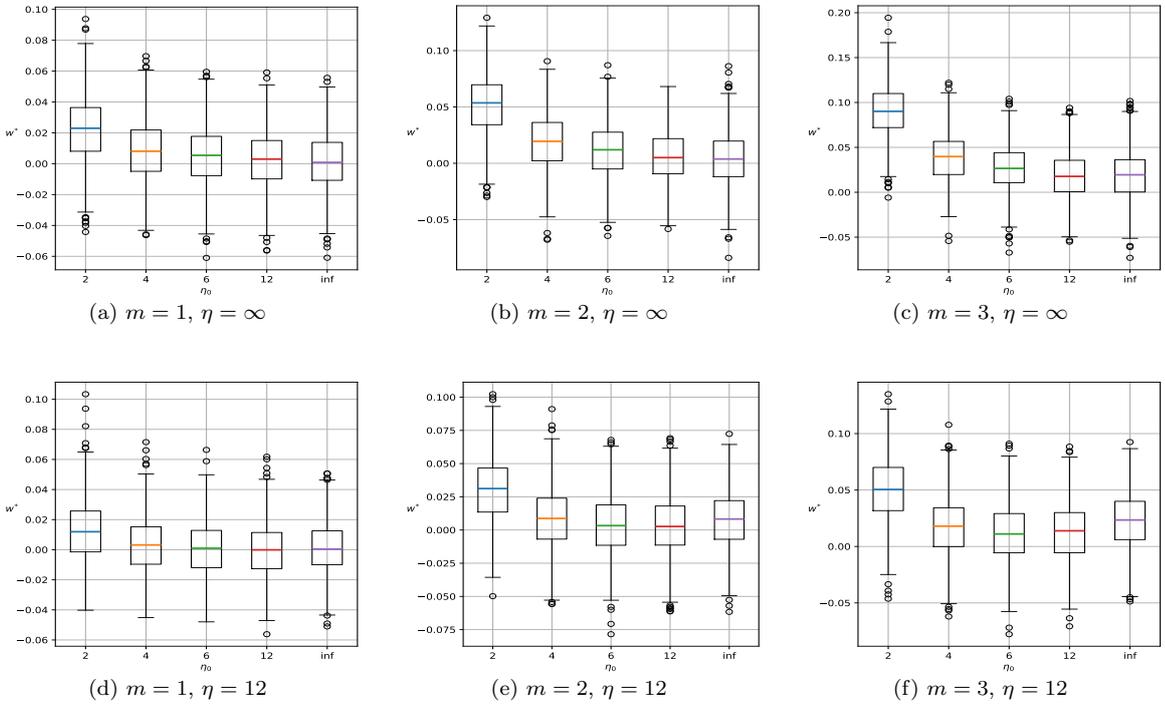

(a) $m=1$, $\eta=\infty$    (b) $m=2$, $\eta=\infty$    (c) $m=3$, $\eta=\infty$

(d) $m=1$, $\eta=12$    (e) $m=2$, $\eta=12$    (f) $m=3$, $\eta=12$

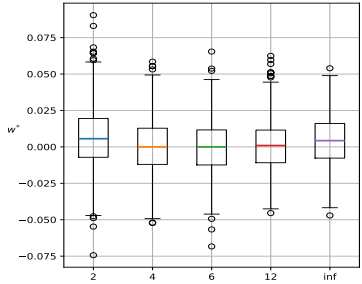
(g) $m = 1$, $\eta = 6$

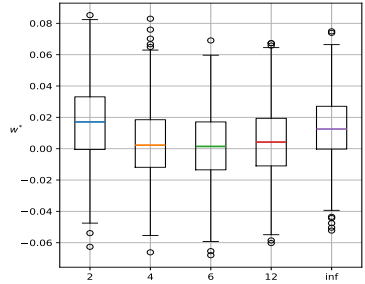
(h) $m = 2$, $\eta = 6$

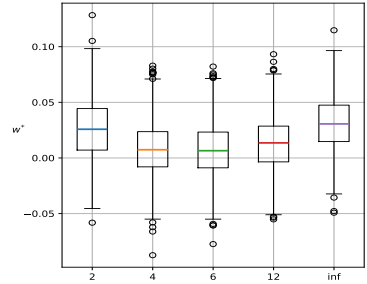
(i) $m = 3$, $\eta = 6$

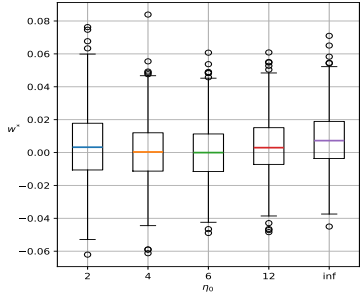
(j) $m = 1$, $\eta = 4$

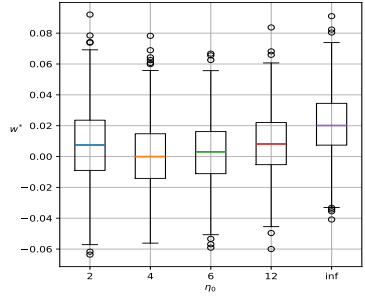
(k) $m = 2$, $\eta = 4$

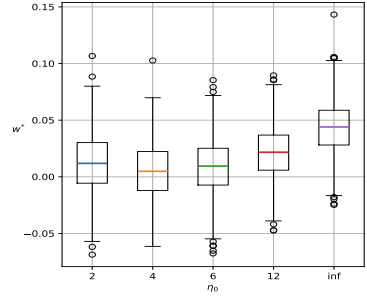
(l) $m = 3$, $\eta = 4$

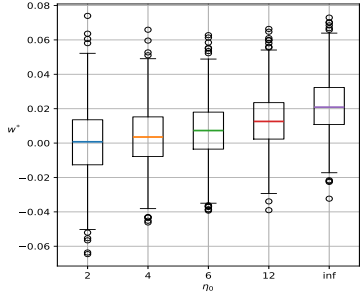
(m) $m = 1$, $\eta = 2$

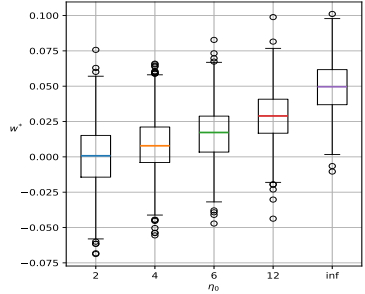
(n) $m = 2$, $\eta = 2$

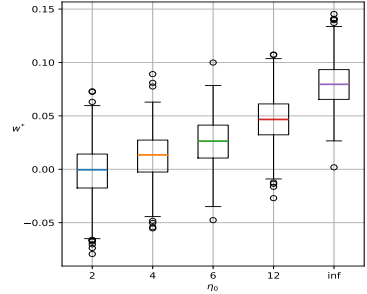
(o) $m = 3$, $\eta = 2$

### 5.3.4 Rates of convergence

As in section 5.2.4, we investigate the rate of convergence of $W_{N,k}^*(m, \eta_0)$ on samples from $P_m(\eta)$ as the sample size $N \to \infty$ by performing linear regression on the model $\log W_{N,k}^*(m, \eta) = \log a + b \log N$.

Table 4 shows the least-squares estimates of the gradients, which serve as estimates for the rates of convergence. As indicated in Figures 8 and 9, when either $\eta$ or $\eta_0$ is small, the test statistic $W_{N,k}^*(m, \eta_0)$ evaluated on samples from $P_m(\eta)$ converges very quickly, and consequently the rates shown in Table 4 are subject to large error unless $\nu$ and $\nu_0$ are both are relatively large, which corresponds to the lower-right portion of each sub-table.

Table 4: Rates of convergence of $W_{N,k}^*(m, \eta_0)$ on samples from $P_m(\eta)$ for $k = 3$

| $m$ | $\eta_0 \backslash \eta$ | 2 | 3 | 4 | 6 | 8 | 11 | 15 | 20 | $\infty$ |
|---|---|---|---|---|---|---|---|---|---|---|
| 1 | 2 | 0.07 | 0.65 | -0.01 | -0.13 | -0.13 | -0.05 | -0.07 | -0.08 | -0.04 |
|  | 3 | -0.14 | -0.35 | -0.45 | -0.23 | -0.30 | -0.19 | -0.15 | -0.19 | -0.12 |
|  | 4 | -0.27 | -0.60 | -0.66 | -0.74 | -0.62 | -0.27 | -0.39 | -0.24 | -0.23 |
|  | 6 | 0.14 | -0.17 | -0.85 | -0.59 | -0.38 | -0.61 | -0.55 | -0.53 | -0.42 |
|  | 8 | 0.08 | -0.07 | -0.23 | -0.68 | -0.76 | -0.94 | -1.08 | -0.75 | -0.52 |
|  | 11 | 0.18 | 0.50 | 0.07 | -0.16 | -0.52 | -0.42 | -0.50 | -0.47 | -0.49 |
|  | 15 | 0.44 | 0.14 | 0.30 | 0.10 | -0.57 | -0.34 | -0.52 | -0.73 | -0.81 |
|  | 20 | 0.24 | 0.21 | 0.04 | -0.02 | -0.50 | -0.67 | -0.56 | -0.69 | -0.73 |
|  | $\infty$ | 0.06 | 0.03 | 0.13 | -0.05 | -0.27 | -0.59 | -0.16 | -0.69 | -0.72 |

Table 4: Rates of convergence of $W^*_{N,k}(m, \eta_0)$ on samples from $P_m(\eta)$ for $k = 3$

| $m$ | $\eta_0$ \ $\eta$ | 2 | 3 | 4 | 6 | 8 | 11 | 15 | 20 | $\infty$ |
|---|---|---|---|---|---|---|---|---|---|---|
| 2 | 2 | -1.22 | 0.06 | 0.04 | -0.06 | -0.05 | -0.07 | -0.08 | -0.06 | -0.05 |
|   | 3 | 0.06 | -0.12 | -0.18 | -0.19 | -0.15 | -0.18 | -0.16 | -0.16 | -0.13 |
|   | 4 | 0.36 | 0.10 | -0.37 | -0.32 | -0.28 | -0.25 | -0.24 | -0.22 | -0.21 |
|   | 6 | -0.09 | -0.28 | -0.50 | -0.69 | -0.66 | -0.51 | -0.45 | -0.46 | -0.35 |
|   | 8 | -0.08 | -0.24 | -0.33 | -0.55 | -0.61 | -0.55 | -0.55 | -0.49 | -0.43 |
|   | 11 | 0.04 | -0.11 | -0.30 | -0.49 | -0.57 | -0.57 | -0.55 | -0.53 | -0.46 |
|   | 15 | -0.01 | -0.08 | -0.21 | -0.34 | -0.44 | -0.51 | -0.46 | -0.49 | -0.48 |
|   | 20 | -0.01 | -0.15 | -0.24 | -0.39 | -0.54 | -0.59 | -0.60 | -0.63 | -0.55 |
|   | $\infty$ | 0.03 | -0.07 | -0.14 | -0.30 | -0.40 | -0.53 | -0.55 | -0.58 | -0.57 |
| 3 | 2 | -0.90 | -0.30 | 0.12 | 0.70 | 0.27 | 0.22 | 0.12 | 0.10 | 0.07 |
|   | 3 | -0.45 | -0.45 | 0.11 | 0.28 | 0.39 | 0.10 | 0.03 | 0.03 | -0.01 |
|   | 4 | 0.38 | 0.15 | -0.07 | 0.42 | 0.03 | -0.05 | -0.04 | -0.04 | -0.07 |
|   | 6 | 0.63 | 0.32 | 0.11 | -0.15 | -0.21 | -0.17 | -0.16 | -0.16 | -0.16 |
|   | 8 | 0.23 | 0.20 | -0.05 | -0.18 | -0.23 | -0.18 | -0.28 | -0.22 | -0.23 |
|   | 11 | 0.10 | 0.07 | -0.07 | -0.14 | -0.23 | -0.27 | -0.28 | -0.28 | -0.26 |
|   | 15 | 0.16 | 0.15 | -0.01 | -0.13 | -0.17 | -0.21 | -0.24 | -0.23 | -0.26 |
|   | 20 | 0.16 | 0.08 | -0.02 | -0.15 | -0.20 | -0.21 | -0.25 | -0.25 | -0.26 |
|   | $\infty$ | 0.09 | 0.01 | -0.05 | -0.14 | -0.18 | -0.24 | -0.25 | -0.28 | -0.29 |

### 5.3.5 Critical values

Table 5 shows estimated critical values $\hat{w}^*_\alpha$ for the distribution of $W^*_{N,k}(m, \eta)$ at significance level $\alpha = 0.05$ with $m \in \{1, 2, 3\}$ $k \in \{1, 2, 3\}$ and $\eta \in \{3, 4, 5, 10, \infty\}$, for sample sizes of $N \in \{100, 200, \ldots, 900, 1000, 2000, \ldots, 5000\}$.

Table 5 again indicates that critical values increase as $m$ increases, and decrease as $\eta$ and $N$ increase, which was also the case Table 2. In addition, Table 5 also shows that critical values tend to *decrease* as $k$ increases, which may be explained by the fact that while the variance of the estimator decreases with $k$.

Table 5: Extimated critical values of $W^*_{N,k}(m, \eta)$ at $\alpha = 0.05$.

| | | $m = 1$ | | | $m = 2$ | | | $m = 3$ | | |
|---|---|---|---|---|---|---|---|---|---|---|
| $\nu$ | $N$ | $k=1$ | $k=2$ | $k=3$ | $k=1$ | $k=2$ | $k=3$ | $k=1$ | $k=2$ | $k=3$ |
| 2 | 100 | 0.661 | 0.275 | 0.231 | 0.576 | 0.296 | 0.233 | 0.641 | 0.296 | 0.230 |
|   | 200 | 0.473 | 0.213 | 0.171 | 0.445 | 0.215 | 0.192 | 0.492 | 0.212 | 0.167 |
|   | 300 | 0.394 | 0.171 | 0.124 | 0.404 | 0.182 | 0.153 | 0.440 | 0.199 | 0.155 |
|   | 400 | 0.355 | 0.141 | 0.112 | 0.358 | 0.150 | 0.117 | 0.348 | 0.154 | 0.136 |
|   | 500 | 0.342 | 0.124 | 0.095 | 0.347 | 0.144 | 0.108 | 0.361 | 0.138 | 0.115 |
|   | 600 | 0.307 | 0.120 | 0.096 | 0.284 | 0.116 | 0.103 | 0.302 | 0.130 | 0.111 |
|   | 700 | 0.247 | 0.111 | 0.089 | 0.313 | 0.118 | 0.096 | 0.286 | 0.126 | 0.106 |
|   | 800 | 0.258 | 0.104 | 0.082 | 0.264 | 0.112 | 0.089 | 0.284 | 0.118 | 0.094 |
|   | 900 | 0.291 | 0.094 | 0.073 | 0.245 | 0.099 | 0.081 | 0.282 | 0.121 | 0.093 |
|   | 1000 | 0.223 | 0.095 | 0.074 | 0.247 | 0.097 | 0.086 | 0.268 | 0.100 | 0.084 |
|   | 2000 | 0.180 | 0.064 | 0.045 | 0.198 | 0.071 | 0.060 | 0.202 | 0.068 | 0.057 |
|   | 3000 | 0.131 | 0.054 | 0.040 | 0.147 | 0.058 | 0.044 | 0.162 | 0.060 | 0.052 |
|   | 4000 | 0.122 | 0.049 | 0.038 | 0.122 | 0.050 | 0.041 | 0.124 | 0.055 | 0.045 |
|   | 5000 | 0.128 | 0.042 | 0.032 | 0.120 | 0.044 | 0.036 | 0.127 | 0.046 | 0.039 |

Table 5: Critical values of $W^*_{N,k}(m,\eta)$ at $\alpha = 0.05$.

| $\nu$ | $N$ | $m=1$ | | | $m=2$ | | | $m=3$ | | |
|---|---|---|---|---|---|---|---|---|---|---|
| | | $k=1$ | $k=2$ | $k=3$ | $k=1$ | $k=2$ | $k=3$ | $k=1$ | $k=2$ | $k=3$ |
| 4 | 100 | 0.350 | 0.233 | 0.219 | 0.435 | 0.277 | 0.261 | 0.459 | 0.319 | 0.275 |
| | 200 | 0.288 | 0.176 | 0.157 | 0.332 | 0.218 | 0.189 | 0.310 | 0.247 | 0.218 |
| | 300 | 0.222 | 0.142 | 0.123 | 0.235 | 0.176 | 0.158 | 0.231 | 0.176 | 0.160 |
| | 400 | 0.195 | 0.124 | 0.111 | 0.214 | 0.137 | 0.125 | 0.236 | 0.169 | 0.155 |
| | 500 | 0.173 | 0.113 | 0.101 | 0.190 | 0.129 | 0.117 | 0.203 | 0.136 | 0.131 |
| | 600 | 0.157 | 0.099 | 0.087 | 0.174 | 0.106 | 0.092 | 0.177 | 0.136 | 0.120 |
| | 700 | 0.146 | 0.094 | 0.076 | 0.157 | 0.113 | 0.100 | 0.169 | 0.122 | 0.111 |
| | 800 | 0.134 | 0.086 | 0.072 | 0.147 | 0.104 | 0.091 | 0.155 | 0.112 | 0.104 |
| | 900 | 0.127 | 0.078 | 0.067 | 0.143 | 0.093 | 0.082 | 0.157 | 0.107 | 0.100 |
| | 1000 | 0.133 | 0.081 | 0.064 | 0.134 | 0.089 | 0.077 | 0.139 | 0.096 | 0.091 |
| | 2000 | 0.088 | 0.052 | 0.046 | 0.092 | 0.063 | 0.056 | 0.095 | 0.076 | 0.066 |
| | 3000 | 0.068 | 0.043 | 0.036 | 0.073 | 0.048 | 0.044 | 0.083 | 0.059 | 0.055 |
| | 4000 | 0.060 | 0.039 | 0.032 | 0.062 | 0.044 | 0.038 | 0.075 | 0.052 | 0.048 |
| | 5000 | 0.058 | 0.036 | 0.029 | 0.060 | 0.038 | 0.034 | 0.063 | 0.047 | 0.044 |
| 12 | 100 | 0.318 | 0.248 | 0.214 | 0.389 | 0.306 | 0.301 | 0.388 | 0.321 | 0.298 |
| | 200 | 0.215 | 0.154 | 0.140 | 0.247 | 0.210 | 0.197 | 0.292 | 0.243 | 0.229 |
| | 300 | 0.174 | 0.134 | 0.115 | 0.218 | 0.171 | 0.170 | 0.251 | 0.204 | 0.198 |
| | 400 | 0.170 | 0.118 | 0.107 | 0.188 | 0.152 | 0.148 | 0.208 | 0.180 | 0.170 |
| | 500 | 0.144 | 0.107 | 0.099 | 0.161 | 0.134 | 0.129 | 0.184 | 0.171 | 0.170 |
| | 600 | 0.133 | 0.096 | 0.088 | 0.147 | 0.117 | 0.117 | 0.166 | 0.149 | 0.148 |
| | 700 | 0.121 | 0.089 | 0.077 | 0.134 | 0.113 | 0.112 | 0.162 | 0.142 | 0.138 |
| | 800 | 0.105 | 0.079 | 0.068 | 0.126 | 0.105 | 0.098 | 0.153 | 0.136 | 0.130 |
| | 900 | 0.108 | 0.079 | 0.069 | 0.124 | 0.104 | 0.094 | 0.132 | 0.117 | 0.113 |
| | 1000 | 0.102 | 0.072 | 0.067 | 0.112 | 0.094 | 0.089 | 0.133 | 0.115 | 0.114 |
| | 2000 | 0.067 | 0.050 | 0.048 | 0.081 | 0.065 | 0.064 | 0.100 | 0.086 | 0.088 |
| | 3000 | 0.055 | 0.042 | 0.038 | 0.062 | 0.051 | 0.048 | 0.083 | 0.071 | 0.071 |
| | 4000 | 0.048 | 0.038 | 0.034 | 0.058 | 0.048 | 0.046 | 0.066 | 0.059 | 0.059 |
| | 5000 | 0.044 | 0.032 | 0.030 | 0.051 | 0.042 | 0.037 | 0.062 | 0.053 | 0.053 |
| 20 | 100 | 0.301 | 0.217 | 0.209 | 0.374 | 0.293 | 0.290 | 0.404 | 0.345 | 0.352 |
| | 200 | 0.215 | 0.159 | 0.150 | 0.245 | 0.217 | 0.210 | 0.282 | 0.234 | 0.236 |
| | 300 | 0.179 | 0.133 | 0.120 | 0.217 | 0.180 | 0.174 | 0.252 | 0.225 | 0.219 |
| | 400 | 0.150 | 0.124 | 0.108 | 0.180 | 0.154 | 0.140 | 0.212 | 0.201 | 0.188 |
| | 500 | 0.133 | 0.101 | 0.093 | 0.170 | 0.149 | 0.131 | 0.196 | 0.179 | 0.172 |
| | 600 | 0.128 | 0.093 | 0.086 | 0.150 | 0.131 | 0.126 | 0.164 | 0.155 | 0.155 |
| | 700 | 0.110 | 0.087 | 0.078 | 0.142 | 0.119 | 0.108 | 0.171 | 0.149 | 0.147 |
| | 800 | 0.103 | 0.079 | 0.075 | 0.132 | 0.107 | 0.106 | 0.157 | 0.141 | 0.144 |
| | 900 | 0.110 | 0.083 | 0.068 | 0.119 | 0.106 | 0.102 | 0.142 | 0.131 | 0.129 |
| | 1000 | 0.100 | 0.073 | 0.062 | 0.119 | 0.095 | 0.090 | 0.132 | 0.120 | 0.122 |
| | 2000 | 0.070 | 0.050 | 0.048 | 0.077 | 0.066 | 0.066 | 0.095 | 0.089 | 0.093 |
| | 3000 | 0.054 | 0.042 | 0.036 | 0.065 | 0.057 | 0.052 | 0.079 | 0.073 | 0.073 |
| | 4000 | 0.048 | 0.038 | 0.034 | 0.057 | 0.050 | 0.046 | 0.069 | 0.063 | 0.062 |
| | 5000 | 0.041 | 0.033 | 0.029 | 0.051 | 0.044 | 0.041 | 0.063 | 0.059 | 0.056 |
| $\infty$ | 100 | 0.293 | 0.239 | 0.223 | 0.346 | 0.309 | 0.293 | 0.417 | 0.390 | 0.380 |
| | 200 | 0.203 | 0.165 | 0.153 | 0.261 | 0.228 | 0.212 | 0.312 | 0.283 | 0.275 |
| | 300 | 0.171 | 0.139 | 0.124 | 0.224 | 0.197 | 0.189 | 0.249 | 0.236 | 0.234 |
| | 400 | 0.149 | 0.115 | 0.101 | 0.180 | 0.164 | 0.165 | 0.223 | 0.208 | 0.211 |
| | 500 | 0.120 | 0.099 | 0.099 | 0.151 | 0.140 | 0.131 | 0.193 | 0.179 | 0.176 |
| | 600 | 0.118 | 0.097 | 0.086 | 0.151 | 0.130 | 0.130 | 0.182 | 0.163 | 0.163 |
| | 700 | 0.116 | 0.089 | 0.077 | 0.131 | 0.115 | 0.111 | 0.170 | 0.151 | 0.153 |
| | 800 | 0.100 | 0.080 | 0.072 | 0.138 | 0.112 | 0.112 | 0.160 | 0.146 | 0.149 |
| | 900 | 0.098 | 0.079 | 0.071 | 0.117 | 0.110 | 0.105 | 0.161 | 0.151 | 0.152 |
| | 1000 | 0.096 | 0.071 | 0.067 | 0.112 | 0.104 | 0.097 | 0.139 | 0.127 | 0.128 |
| | 2000 | 0.064 | 0.046 | 0.047 | 0.081 | 0.071 | 0.071 | 0.096 | 0.098 | 0.101 |
| | 3000 | 0.053 | 0.042 | 0.039 | 0.065 | 0.057 | 0.058 | 0.084 | 0.079 | 0.081 |
| | 4000 | 0.049 | 0.037 | 0.034 | 0.052 | 0.047 | 0.046 | 0.069 | 0.069 | 0.071 |
| | 5000 | 0.040 | 0.032 | 0.029 | 0.052 | 0.044 | 0.043 | 0.060 | 0.061 | 0.062 |

### 5.3.6 Statistical power

For the null hypothesis $\eta = \eta_0$, the power of the test to detect the alternative hypothesis $\eta \neq \eta_0$ at significance level $\alpha$ is defined by
$$\gamma(\eta) = \mathbb{P}_\eta\big(W^*_{N,k}(m,\eta_0) > w^*_\alpha\big), \qquad \nu \neq \nu_0,$$
where $w^*_\alpha$ is the $(1-\alpha)$-quantile of the distribution of $W^*_{N,k}(m,\eta_0)$ under the null hypothesis, and $\mathbb{P}_\eta$ is defined by distribution of $W_{N,k}(m,\eta_0)$ on samples from the $P_m(\eta)$ distribution.

We estimate $\gamma(\eta)$ using the estimated critical values $\hat{w}^*_\alpha$ computed in the previous section: let $\{w^*_1, w^*_2, \ldots, w^*_M\}$ independent observations of the test statistic $W^*_{N,k}(m,\nu_0)$ on independent samples from the $P_m(\eta)$ distribution. As in section 5.2.6 we estimate the power function by the estimator
$$\hat{\gamma}(\eta) = \frac{1}{M} \sum_{j=1}^{M} I(w^*_j > \hat{w}^*_\alpha)$$
where $I$ is the indicator function. Table 6 shows the estimated power for the case $N=5000$ and $k=3$ at significance level $\alpha = 0.05$.

Table 6: Estimated power of $W^*_{N,k}(m,\eta_0)$ on $P_m(\eta)$
for $N = 5000, k = 3, \alpha = 0.05$

| $m$ | $\eta_0 \backslash \eta$ | 2 | 3 | 4 | 6 | 8 | 10 | 12 | 15 | 20 | $\infty$ |
|---|---|---|---|---|---|---|---|---|---|---|---|
| 1 | 2 | 0.05 | 0.05 | 0.05 | 0.07 | 0.09 | 0.11 | 0.10 | 0.14 | 0.14 | 0.25 |
|   | 3 | 0.08 | 0.05 | 0.05 | 0.06 | 0.06 | 0.07 | 0.07 | 0.09 | 0.10 | 0.15 |
|   | 4 | 0.11 | 0.06 | 0.05 | 0.04 | 0.06 | 0.05 | 0.07 | 0.07 | 0.08 | 0.08 |
|   | 6 | 0.15 | 0.08 | 0.06 | 0.05 | 0.05 | 0.04 | 0.05 | 0.06 | 0.06 | 0.09 |
|   | 8 | 0.17 | 0.08 | 0.06 | 0.05 | 0.05 | 0.04 | 0.04 | 0.05 | 0.04 | 0.06 |
|   | 10 | 0.20 | 0.11 | 0.08 | 0.05 | 0.05 | 0.05 | 0.05 | 0.05 | 0.04 | 0.05 |
|   | 12 | 0.19 | 0.11 | 0.08 | 0.06 | 0.05 | 0.05 | 0.05 | 0.05 | 0.04 | 0.06 |
|   | 15 | 0.21 | 0.11 | 0.08 | 0.07 | 0.06 | 0.05 | 0.05 | 0.05 | 0.04 | 0.05 |
|   | 20 | 0.26 | 0.16 | 0.10 | 0.07 | 0.07 | 0.06 | 0.05 | 0.05 | 0.05 | 0.06 |
|   | $\infty$ | 0.36 | 0.21 | 0.13 | 0.08 | 0.08 | 0.07 | 0.06 | 0.07 | 0.05 | 0.05 |
| 2 | 2 | 0.05 | 0.06 | 0.08 | 0.15 | 0.24 | 0.29 | 0.37 | 0.42 | 0.49 | 0.77 |
|   | 3 | 0.08 | 0.05 | 0.05 | 0.06 | 0.11 | 0.13 | 0.14 | 0.19 | 0.21 | 0.39 |
|   | 4 | 0.12 | 0.07 | 0.05 | 0.08 | 0.08 | 0.08 | 0.11 | 0.13 | 0.15 | 0.26 |
|   | 6 | 0.21 | 0.10 | 0.06 | 0.05 | 0.05 | 0.07 | 0.06 | 0.08 | 0.08 | 0.12 |
|   | 8 | 0.28 | 0.12 | 0.07 | 0.07 | 0.05 | 0.06 | 0.06 | 0.07 | 0.06 | 0.11 |
|   | 10 | 0.33 | 0.14 | 0.08 | 0.05 | 0.05 | 0.05 | 0.05 | 0.05 | 0.07 | 0.08 |
|   | 12 | 0.40 | 0.17 | 0.11 | 0.08 | 0.06 | 0.07 | 0.05 | 0.06 | 0.06 | 0.08 |
|   | 15 | 0.38 | 0.13 | 0.10 | 0.05 | 0.04 | 0.04 | 0.05 | 0.05 | 0.05 | 0.05 |
|   | 20 | 0.45 | 0.20 | 0.12 | 0.07 | 0.06 | 0.06 | 0.05 | 0.06 | 0.05 | 0.06 |
|   | $\infty$ | 0.64 | 0.30 | 0.19 | 0.10 | 0.08 | 0.06 | 0.05 | 0.05 | 0.05 | 0.05 |
| 3 | 2 | 0.05 | 0.07 | 0.11 | 0.28 | 0.44 | 0.53 | 0.65 | 0.71 | 0.81 | 0.98 |
|   | 3 | 0.07 | 0.05 | 0.07 | 0.10 | 0.17 | 0.25 | 0.30 | 0.35 | 0.43 | 0.76 |
|   | 4 | 0.11 | 0.06 | 0.05 | 0.07 | 0.09 | 0.12 | 0.17 | 0.19 | 0.24 | 0.50 |
|   | 6 | 0.23 | 0.09 | 0.05 | 0.05 | 0.06 | 0.07 | 0.07 | 0.10 | 0.14 | 0.26 |
|   | 8 | 0.32 | 0.11 | 0.05 | 0.04 | 0.05 | 0.06 | 0.07 | 0.07 | 0.09 | 0.15 |
|   | 10 | 0.40 | 0.14 | 0.09 | 0.06 | 0.05 | 0.05 | 0.06 | 0.08 | 0.07 | 0.15 |
|   | 12 | 0.46 | 0.18 | 0.07 | 0.05 | 0.04 | 0.05 | 0.05 | 0.06 | 0.05 | 0.11 |
|   | 15 | 0.57 | 0.19 | 0.12 | 0.06 | 0.06 | 0.06 | 0.05 | 0.05 | 0.06 | 0.10 |
|   | 20 | 0.59 | 0.21 | 0.13 | 0.06 | 0.06 | 0.05 | 0.04 | 0.04 | 0.05 | 0.08 |
|   | $\infty$ | 0.84 | 0.41 | 0.19 | 0.09 | 0.06 | 0.05 | 0.05 | 0.04 | 0.05 | 0.05 |

Table 6 indicates that the power to detect $\eta \neq \eta_0$ increases as the dimension $m$ increases, and also increases as the difference $|\eta - \eta_0|$ increases. Overall the power of the test is low compared with that of the test based on $W_{N,k}(m,\nu_0)$, which is perhaps because the sample variance of $W^*_{N,k}(m,\eta_0)$ is relatively large compared to the differences between its limiting values for different values of $\eta \neq \eta_0$.

## 5.4 Conclusion

We have proposed goodness-of-fit statistics for the multivariate Student and multivariate Pearson type II distributions, based on the maximum entropy principle and a class of estimators for Rényi entropy based on nearest neighbour distances. We have proved the $L^2$-consistency of these statistics using results on the subadditivity of Euclidean functionals on nearest neighbour graphs, and investigated their distributions and rates of convergence using Monte Carlo methods. Our numerical results indicate that hypothesis tests based on these statistics behave as one might expect, however for samples of up to 5000 points their statistical power rather disappointing, especially for samples from the Pearson type II distribution, which is probably due to the fact that the variance of our test statistics is relatively large compared to the size of the effect we seek to identify.

**Acknowledgements**

Nikolai Leonenko (NL) would like to thank for support and hospitality during the programme "Fractional Differential Equations" and the programmes "Uncertainly Quantification and Modelling of Material" and "Stochastic systems for anomalous diffusion" in Isaac Newton Institute for Mathematical Sciences, Cambridge. These programmes were organized with the support of the Clay Mathematics Institute, of EPSRC (via grants EP/W006227/1 and EP/W00657X/1), of UCL (via the MAPS Visiting Fellowship scheme) and of the Heilbronn Institute for Mathematical Research (for the Sci-Art Contest). Also NL was partially supported under the ARC Discovery Grant DP220101680 (Australia), LMS grant 42997 (UK), Croatian Scientific Foundation (HRZZ) grant "Scaling in Stochastic Models" (IP-2022-10-8081) and grant FAPESP 22/09201-8 (Brazil). Vitali Makogin was supported by the Grant No. 39087941 of the German Research Society.